# Localized and delocalized states of a diamine cation: A critical test of wave function methods

*Marta Gałyńska[a], Vilhjálmur Ásgeirsson[a], Hannes Jónsson[a]\*, Ragnar Bjornsson[b]\**

[a] Faculty of Physical Science, VR-III, University of Iceland, 107 Reykjavik, Iceland

[b] Max-Planck Institute for Chemical Energy Conversion, Stiftstrasse 34-36, 45470 Mülheim an der Ruhr, Germany.

\***Corresponding author**: ragnar.bjornsson@cec.mpg.de, hj@hi.is

**ABSTRACT**

The relative stability of a localized and delocalized electronic state in the same molecule, the N,N'-Dimethylpiperazine cation, is calculated at various levels of theory up to multireference configuration interaction (MRCI+Q). This system has received a great deal of attention because of recent experimental studies of corresponding Rydberg states of the molecule and the failure of most density functional approximations to produce a metastable localized state. A cut through the energy surface involving two dihedral angles is generated at the level of MRCI+Q as well as Hartree-Fock (HF), Möller-Plesset second order perturbation theory (MP2), coupled cluster theory with and without perturbative triple excitations (CCSD and CCSD(T)) and complete active space self-consistent field calculations with and without perturbative correction (CASSCF and NEVPT2). Remarkably, while CCSD produces a localized state, CCSD(T) does not, and similarly, large active-space CASSCF does while NEVPT2 does not. The inclusion of dynamic correlation in a perturbative way thus adversely affects the accuracy of the calculation, most notably for CCSD(T), the 'golden standard'. The MRCI+Q results are in close correspondence with the experimental results, as well as CAS(19,20) DMRG-CASSCF calculations. The results presented here establish a benchmark system for the study of electronic state localization.

1. INTRODUCTION

Charge transfer (CT) is a fundamental but rather simple reaction, which does not involve any breaking or creation of bonds, but only migration of a charge within a system. Intervalence charge transfer involves at least two redox centers, playing the role of donor and acceptor sites, and these sites can be on the same molecular fragment or different ones. The charge can move between those two centers directly (through-space) or through saturated or unsaturated bridges (through-bond). Such a system usually belongs to the class of mixed-valence systems, which are defined as compounds containing two or more identical redox centers but in different oxidation states.[1] Molecular mixed-valence systems are typically open-shell transition metal compounds,[2] but can also be organic[3,4] and they often play an integral role in redox- and photocatalysis[5,6] as well as electron transfer in biological systems.[7,8] The study of simple molecular model systems, can provide useful information about the mechanism and rates of charge separation, delocalization, and charge transfer from donor to acceptor center.

N,N'-dimethylpiperazine (DMP) was identified as an interesting organic model system to study intramolecular charge transfer via through-bond and through-space interactions.[9,10,11,12,13,14,15] This simple six-membered ring contains two identical ionization centers (nitrogen atoms) connected via saturated C-C bridges, where the charge can either localize on one of the nitrogen atoms or delocalize between the two nitrogen atoms. Experimental studies on $DMP_+$ were originally performed in solution;[9,10,11] while the EPR spectra were not conclusive with respect to the electronic structure being localized or delocalized[11], a comparison between resonance Raman spectra and computed vibrational frequencies found an electronic structure most consistent with a delocalized structure.[9,10] Until recently, a localized electronic structure of $DMP_+$, had not been observed experimentally. Using ultrafast time-resolved Rydberg spectroscopy in the gas phase where an excitation from the ground state of the DMP molecule to the 3p Rydberg state allowed one to monitor the picosecond time-scale dynamics in going from a localized to a delocalized cation state and to measure the energy difference (0.33 eV, in favour of the delocalized state)[12,13], a rare observation for a mixed-valence system. It is assumed in these studies that the Rydberg states involved, closely resemble the cationic electronic states.[13]

In a recent joint experimental-theoretical study[13] of DMP+, it was shown that many common density functional methods failed to find the localized state of DMP+ and that only methods that reduce self-interaction error sufficiently, were capable of giving a result in agreement with experiment. This has subsequently led to a debate in the literature[14,15] about whether DMP+ really represents a failure of common density functional approximations, the height or even existence of the barrier between the delocalized and localized state at various levels of theory and whether coupled cluster theory is an appropriate high-level reference theory to compare DFT results to.

Organic mixed-valence cations such as DMP+, have actually been found to present considerable challenges to both state-of-the-art wavefunction theory and density functional theory approaches.[16,17] As discussed by Kaupp and coworkers, wavefunction theory approaches based on unrestricted HF wavefunctions have an initial bias towards symmetry breaking and hence charge localization, in addition to spin contamination. Electron correlation is thus crucial to describe the possible delocalization present in such systems, but unrestricted MP2 calculations have been found to suffer from exaggerated spin contamination. Therefore, a robust dynamic correlation treatment appears necessary. Furthermore, as these systems have near degeneracies, it may be important to describe the static correlation reliably from the start (e.g. via multiconfigurational approaches) in addition to the dynamic correlation (though these phenomena are not always clearly separable).

In density functional theory, the long-standing issue of self-interaction/delocalization error and the amount of HF exchange in hybrid functionals becomes a critical issue for mixed-valence systems. Regular semilocal functionals and common hybrid functionals tend to prefer an overdelocalized description of such systems, likely due to self-interaction and delocalization errors.[18,19,20,21] A common way of dealing with self-interaction/delocalization errors is to increase the amount of Hartree-Fock exchange in the hybrid DFT functional form. In this way, it was found that hybrid functionals with higher HF exchange e.g. >35 % are necessary in some cases for describing correctly molecules with more localized charge.[16] However, too much HF exchange, like in the HF method, may lead to an erroneously localized description of delocalized systems. Kaupp and coworkers have discussed several inorganic[22,23] and organic[17,24,25,26,27] mixed-valence systems, that presents considerable challenges for DFT approaches. Benchmarking of both DFT and WFT approaches with respect to highly accurate CCSDT(Q)/CBS results has sometimes been possible owing to the small size. It has been found that for global hybrids with 35-43% HF exchange, range-separated hybrids or a local hybrid functional were capable of achieving good agreement with the high-level theory.[17,22,24,25,26,27] Functionals with too little HF exchange gave an erroneous delocalized description while too much HF exchange gave an overlocalized description leading to too large energy differences. Nonempirically tuned range-separated hybrid DFT[28,29,30] has also recently been successfully used to study delocalization in three organic mixed valence molecules.[31] A less common way of dealing with self-interaction/delocalization errors is the use of the Perdew-Zunger[18] (PZ-SIC) self-interaction correction in the DFT functional. This approach has recently been implemented as a variational, self-consistent method based on complex orbitals[32,33] and successfully used to study molecules and solids where self-interaction/delocalization errors play a large role.[34,35,36,37,38]

Multiconfigurational methods are nowadays primarily based on the Complete Active Space SCF (CAS-SCF) approach, that has been previously used to study mixed-valence systems.[39,40, 41,42,43,44,45] In CASSCF a chemically important active space is defined, intended to capture the important static correlation, in which Full-CI is applied. However, this typically limits the active space to 14-16 orbitals. Modern approximations to Full-CI such as the DMRG algorithm[46,47,48] has expanded this active space limit to ~50 orbitals. As CASSCF-based approaches only capture the static correlation, multireference perturbation theory (CASPT2, NEVPT2) and multireference configuration interaction (e.g. MRCI+Q) are typically employed to capture the missing dynamic correlation. An affordable and robust simultaneous treatment of both static and dynamic correlation; however, is still a challenge for contemporary quantum chemistry method development.[49,50]

In this study, we perform an extensive theoretical investigation of the potential energy surface of the DMP$_+$ cation, employing both single-reference and multireference wavefunction theory methods and compare to density functional theory descriptions. We discuss the multireference nature of the DMP cation and discuss how Hartree-Fock qualitatively fails to describe the PES, which in turns affects all single-reference wavefunction methods that use the HF wavefunction as a reference determinant. Multiconfigurational SCF calculations at the CASSCF and DMRG-CASSCF levels and multireference configuration interaction (FIC-MRCI+Q) are used to firmly establish the potential energy surface of the DMP cation and quantitative agreement with experiment is achieved. Severe problems are, however, found when using perturbation theory both in the multi-reference case (NEVPT2) and in the single-reference case (CCSD(T)).

## 2. METHODOLOGY

The DMP cationic potential energy surface was investigated using different levels of theory, based on single- and multireference wavefunction theory (WFT) and density functional theory (DFT). All calculations were performed using the ORCA 4.0 quantum chemistry code.[51] Single-determinant methods used included Hartree-Fock (HF)[52,15,54] and hybrid density functional methods such as B3LYP[55] and BHLYP[56] functionals, containing 20% and 50% of HF exchange, respectively. Calculations were performed using the correlation-consistent basis sets, primarily cc-pVDZ and aug-cc-pVDZ[57,58]. Larger basis sets (aug-cc-pVTZ) were also used for some calculations as discussed later. Coupled cluster calculations were performed at the singles and doubles level (CCSD)[59,60,61,62], as well as including the perturbative triples correction (CCSD(T))[63] using a UHF reference wavefunction if not mentioned (see SI section 7 for other calculations with other references). Saddle points connecting the localized and delocalized minima were located at the BHLYP level of theory using an eigenvector-following method using the exact Hessian as implemented in ORCA. Hessians were calculated analytically for HF and DFT methods. Stability analysis was performed for all HF calculations.[64]

The Complete Active Space Self-Consistent Field (CASSCF) method was used for multireference calculations with the Density Matrix Renormalization Group (DMRG) approximation used for the large active-space CASSCF calculations (active spaces of CASSCF(19,20) etc.)[65,66,67] using the Block code interfaced to ORCA.[68,69,70,71,72] It was shown that the number of renormalized states (details in Table S3 in SI) had a very small influence on the energy difference between the localized and delocalized DMP$_+$ states and all calculations were performed using M=500 renormalized states. Starting orbitals for CASSCF and DMRG-CASSCF calculations were MP2 natural orbitals localized with the Pipek-Mezey method[73] and further rotated to obtain the desired orbitals in the active space. The gradient used for the DMRG-CASSCF optimizations is an approximation to the complete analytical gradient. As the DMRG wavefunction is not invariant to active-space rotations, active-active rotation contributions should be included in the complete gradient,[74] but are omitted here. This results in a small error, which was demonstrated to be negligible in our calculations of DMP$_+$ compared to the analytical CASSCF gradient calculations for smaller active spaces, which yielded near-identical results (see Table S4 in SI). DMRG-CASSCF Hessians were calculated numerically. Several active spaces were used, of different size and composition. The minimal active space of CASSCF(3,2) consists of 3 electrons in the 2 nitrogen lone pair orbitals. A larger active space of CASSCF(7,8) includes in addition the σ H$_3$C-N bonding orbitals and corresponding virtual orbitals. A chemically more reasonable (as will be discussed) active space of same size, CASSCF(7,8)' includes the two σ C-C bonding and corresponding σ* orbitals instead. The active space of CASSCF(11,12) includes both σ H$_3$C-N and C-C orbitals (and corresponding virtual orbitals). The largest active space calculated (via DMRG-CASSCF calculations) comprised of eight σ orbitals (including six C-N and two C-C orbitals) as well as two nonbonding N orbitals, with corresponding virtual orbitals, which was 19 electrons in 20 orbitals, CASSCF(19, 20). All active space orbital sets are shown in Figure S4-S8.

Dynamic correlation in the multireference calculations was described either via N-Electron Valence Perturbation Theory (NEVPT2)[75,76] or by multi-reference configuration interaction (MRCI)[77]. A DMRG-NEVPT2 implementation has recently been described by Chan and coworkers,[76] requiring the 4$_{th}$-order reduced density matrix. The fully internally contracted version of MRCI (FIC-MRCI[78]) including the Davidson size-consistency correction[79] for unlinked quadruples (Q) methods, as implemented in ORCA was used (FIC-MRCI+Q). The FIC-MRCI method utilizes internal contraction that avoids bottlenecks associated with the traditional uncontracted MRCI approaches that generate excited configuration state functions (CSFs) from each reference wavefunction CSF. The FIC-MRCI instead generates excited CSFs by applying an excitation operator to the whole reference wavefunction.

3. RESULTS

We will begin by discussing the molecular and electronic structures of the delocalized and localizes DMP+ states (section 3.1). Next, the cut of the potential energy surface will be introduced (section 3.2) that conveniently relates the localized and delocalized minima (DMP-L+ and DMP-D+), which will be used throughout the subsequent sections. We then begin by presenting the potential energy surfaces of DMP+ at the BHLYP and FIC-MRCI+Q levels of theory (section 3.2.1), and this is then followed by section 3.3 that shows the analogous surfaces using single-reference wavefunction theory methods (HF, MP2, CCSD and CCSD(T)). Section 3.4 then presents the results of multireference wavefunction theory methods (CASSCF, DMRG-CASSCF, NEVPT2, FIC-MRCI+Q).

**3.1 Molecular and electronic structure of DMP+**

The structures of neutral DMP, its two cationic conformers, DMP-L+ and DMP-D+, and the saddle point DMP-SP+ structure connecting the DMP-L+ and DMP-D+ states, relaxed at the BHLYP/aug-cc-pVDZ level of theory are shown in Figure 1.

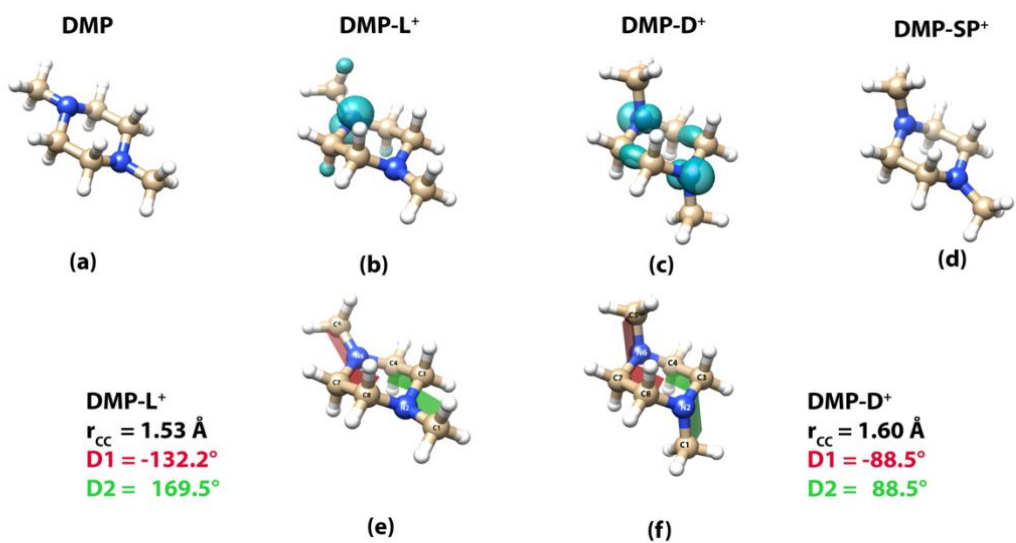

**Figure 1.** Structures of (a) neutral N,N'-Dimethylpiperazine, DMP (the primary eq-eq conformer), (b) and (e) localized cation, DMP-L+, (c) and (f) delocalized cation, DMP-D+, and (d) saddle point for the transition between DMP-L+ and DMP-D+, denoted DMP-SP+. The spin densities shown in (b) and (c) correspond to isosurface level of 0.01 electrons/Å$^3$. The definition of the dihedral angles D1 and D2 that are used to span a cut through the energy surface are shown in (e) and (f), and their values in DMP-L+ and DMP-D+ given, as well as the C-C bond length in the two structures.

The lowest energy structure of the neutral DMP molecule is a chair conformer of $C_{2h}$ symmetry (both methyl groups in equatorial positions) (Figure 1a), which clearly reveals the presence of lone pairs on each sp$_3$-hybridized nitrogen (atoms N2 and N6 in Figure 1). Ionization of alkylamines is known to effectively result in electron removal from nitrogen lone pairs.[80] After removing one electron from a nitrogen lone pair orbital, a positive hole remains, which results in Jahn-Teller-type distortion and the formation of a distorted structure of $C_s$ symmetry. The geometry of this localized structure, DMP-L+, is shown in Figure 1b (a symmetrically equivalent conformer also exists where the hole is localized on the other nitrogen atom). The structure of DMP-L+ clearly reflects the different electronic nature of the nitrogen atoms, where the ionized nitrogen site exhibits more sp$_2$-like character and the non-ionized site remains sp$_3$-like. This is reflected in the spin density in Figure 1b, showing the unpaired electron localized on only one nitrogen atom. Alternatively, the charge can be delocalized between both nitrogen atoms, resulting in a DMP-D+ structure of *C$_{2h}$* symmetry. The delocalized DMP+ structure shown in Figure 1c reveals the spin density of DMP-D+ as delocalized over both nitrogen atoms and interestingly a contribution from the bridging C-C atoms can be seen, suggesting the involvement of the C-C bonds in lone pair interactions that results in the delocalized state.

The charge localization and delocalization in DMP+ involves geometrical changes such as the bending and rotation of the methyl groups, as well as, elongation of C-C bonds. We found that two dihedral angles, D1 and D2, serve the purpose of being suitable descriptors for characterizing the cut of the potential energy surface that connects both delocalized and localized minima. The D1 (D2) angle is defined via the two planes created by C5-N6-C7 (C1-N2-C3) and N6-C7-C8 (N2-C3-C4) atoms (Figure 1b and 1c). These coordinates differ significantly between the DMP-L+ and DMP-D+ structures and lead to a clear separation of the two minima. The D1 and D2 angles as well as the C-C bond lengths for neutral DMP, DMP-D+ and DMP-L+ obtained from structures relaxed with different levels of theory are listed in Table S1 and discussed in detailed in Supporting Information. Interestingly, rather few quantum chemistry methods, including BHLYP, CCSD, CASSCF(11,12), DMRG-CASSCF(19,20) locate both DMP-D+ and DMP-L+ minima in geometry optimizations. The small active-space CASSCF methods ( (3,2), (7,8) and (7,8)' ) as well as HF were not able to locate the delocalized minimum, while DFT functionals with <50 % HF exchange were not able to locate the localized minimum.

**3.2 Potential energy surfaces**

The absence of the DMP-D+ and DMP-L+ minima for some electronic structure methods according to the geometry optimizations, as well as the strong variability in the computed structures creates problems about how to discuss and compare the potential energy surface of DMP+ with different electronic structure methods and brings up the question of what method is really capable of describing the potential energy surface of DMP+ correctly and why methods behave so differently. In order to conveniently compare different methods and to quantify the differences between methods, we have used a 78-point potential energy surface cut using the dihedral angles D1 and D2, which vary from 70° to 175°. The surfaces were interpolated using a biharmonic spline interpolation (v4) provided in Matlab.[81] Constrained geometry optimizations at the BHLYP/aug-cc-pVDZ level were performed where the D1 and D2 angles were fixed while all other coordinates were minimized. Other surfaces using a different electronic structure method were calculated via single point calculations on the constrained optimized BHLYP structures.

The choice to use BHLYP to generate the geometries for all surfaces is justified as BHLYP is one of few methods (practically the only DFT method) that gave converged minima for both DMP-L+ and DMP-D+ and for which analytical gradients are readily available. We also compared the quality of the BHLYP structures to CCSD- and DMRG-CASSCF(19,20) structures for DMP-L+ and DMP-D+ via single-point FIC-MRCI+Q/aug-cc-pVDZ single point calculations. The FIC-MRCI+Q total energies (see Table S2 in the SI) were lowest when using CCSD structures (hence being closer to the FIC-MRCI+Q minima) but the BHLYP structures were not far off and better than the DMRG-CASSCF(19,20) structures. Finally, the energy difference between DMP-L+ and DMP-D+ is hardly affected (< 0.02 eV) by the geometric difference between CCSD and BHLYP as shown in Table 1c.

**Table 1.** Energy differences (eV) of DMP-D+ and DMP-SP+ relative to DMP-L+ calculated using the various relaxed structures as well as the different levels of theory. The total energies [$E_h$] are listed in Table S2 in SI.

| Method | DMP-L+ - DMP-D+ | DMP-L+ - DMP-SP+ |
|---|---|---|
| Experiment | 0.33[12] | ~0.1[15] |
| *a. Relaxed structures at each theory level* | | |
| BHLYP/aug-cc-pVDZ | 0.178 | 0.033 |
| DMRG-CASSCF(19,20)/aug-cc-pVDZ | 0.251 | 0.071 |
| DMRG-CASSCF(19,20)/aug-cc-pVTZ | 0.236 | 0.072 |
| CCSD/aug-cc-pVDZ | 0.23[a] | |
| *b. Single point FIC-MRCI+Q(11,12)/aug-cc-pVDZ calculations on the relaxed minima* | | |
| BHLYP/aug-cc-pVDZ | 0.336 | |
| CCSD/aug-cc-pVDZ | 0.321 | |
| DMRG-CASSCF(19,20)/aug-cc-pVDZ | 0.319 | |
| *c. Single point calculations on the relaxed BHLYP minima* | | |
| MP2/aug-cc-pVDZ | 0.153 | |
| CCSD/aug-cc-pVDZ | 0.221 | |
| OO-CCD/aug-cc-pVDZ | 0.247 | |
| CASSCF(11,12)/aug-cc-pVDZ | -0.017 | |
| DMRG-CASSCF(19,20)/aug-cc-pVDZ | 0.265 | |
| NEVPT2(3,2)/aug-cc-pVDZ | 0.136 | |
| NEVPT2(7,8)/aug-cc-pVDZ | 0.743 | |
| NEVPT2(7,8)'/aug-cc-pVDZ | 0.899 | |
| NEVPT2(11,12)/aug-cc-pVDZ | 0.769 | |

| | |
|---|---|
| FIC-MRCI+Q (3,2)/aug-cc-pVDZ[b] | -0.164 |
| FIC-MRCI+Q (7,8)/aug-cc-pVDZ | 0.261 |
| FIC-MRCI+Q (7,8)'/aug-cc-pVDZ | 0.307 |
| FIC-MRCI+Q(11,12)/aug-cc-pVDZ | 0.336 |

[a] The energy difference was taken from Ref.13, [b] The electronic structure is symmetry-broken for the D+ state.

### 3.2.1 DFT surfaces vs. FIC-MRCI+Q

Before presenting the results of the successes and failures of various single-reference (section 3.3) and multi-reference wavefunction methods (section 3.4) in detail, we will first present the energy surface with BHLYP (which was used to generate the structures used for all surfaces), compare the results to that of a representative density functional (B3LYP) and then show the results obtained with the most accurate wavefunction theory-based approach employed: FIC-MRCI+Q.

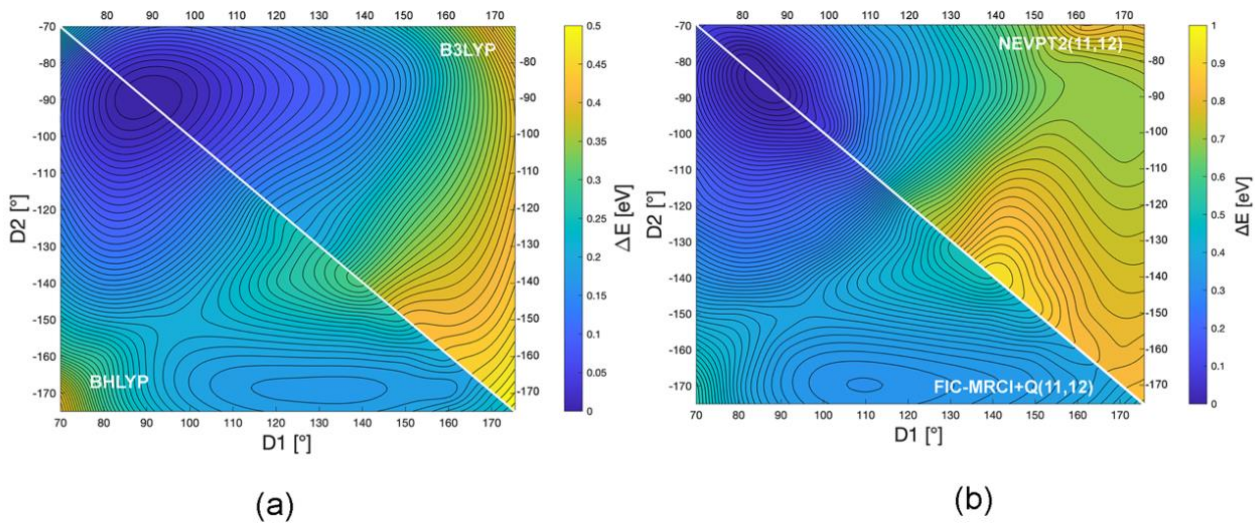

**Figure 2.** The potential energy surfaces calculated at the (a) B3LYP/aug-cc-pVDZ (upper) and BHLYP/aug-cc-pVDZ (lower) and (b) (a) NEVPT2/aug-cc-pVDZ (upper) and FIC-

MRCI+Q(11,12)/cc-pVDZ (lower) level of theory on the constrained-optimized BHLYP structures.

Figure 2a (bottom) shows the energy surfaces with BHLYP. BHLYP gives an energy surface with two well-resolved minima and an apparent saddle-point at D1=-97.3°, D2=151.8°. The energy difference between $D_+$ and $L_+$ was found to be +0.178 eV ($D_+$ more stable), in respectable agreement with the experimental estimate of +0.33 eV. A barrier of +0.033 eV in going from $L_+$ to $D_+$ is found which is also consistent with the experimental estimate of ~0.1 eV. As will be shown, BHLYP is one of very few single-determinant methods that gives an energy surface of DMP+ that is qualitatively consistent with both experiment and high-level quantum chemistry calculations. In sharp contrast to the BHLYP surface, the B3LYP surface shows the complete absence of a localized minimum. This absence of $L_+$ is in agreement with a previous study[13], and exhibits a typical behaviour seen for both non-hybrid functionals as well as hybrid density functionals with <50 % HF exchange. Only a single minimum is found on the surface, which corresponds to the DMP-$D_+$ state in the D1=-90°, D2=90° coordinate region. This finding is clearly inconsistent with the experimental result (where two states separated by a barrier is found) and interestingly B3LYP (and most other functionals) appears to suffer from overstabilization of the DMP-$D_+$ region while failing to describe the DMP-$L_+$ region correctly. This result of common density functionals has been interpreted as arising due to the self-interaction error of DFT where electronic structure is often described as too delocalized.[13]

Shown in Figure 2b (bottom), is the energy surface calculated with the most accurate quantum chemistry method used in this study, the multireference method FIC-MRCI+Q, using a large active space of CAS(11,12). Reassuringly, the energy surfaces of BHLYP and FIC-MRCI+Q are overall very similar, exhibiting resolved minima for both $D_+$ and $L_+$ as well as a clear saddle-point region and impressively, the energy difference is +0.326 eV, in excellent agreement with the experimental estimate of +0.33 eV. As will be discussed, FIC-MRCI+Q is one of relatively few wavefunction-based methods that appears to give a balanced description of electron correlation on the DMP$_+$ potential energy surface. Figure 2b top shows the surprising failure of another multireference method (NEVPT2) that is further discussed in section 3.4.

## 3.3 Single-reference wavefunction surfaces

Single-reference post-HF wavefunction methods have been previously used to describe the DMP cation in the previous papers [9,10,13,14] but without a thorough analysis of the electronic structure and energy surfaces. We will start by describing the HF surface and how electron correlation changes the energy surface for each post-HF method.

### *3.3.1 Hartree-Fock and MP2 energy surfaces*

The potential energy surfaces calculated at the HF/aug-cc-pVDZ and MP2/aug-cc-pVDZ level of theory are shown in Figure 3.

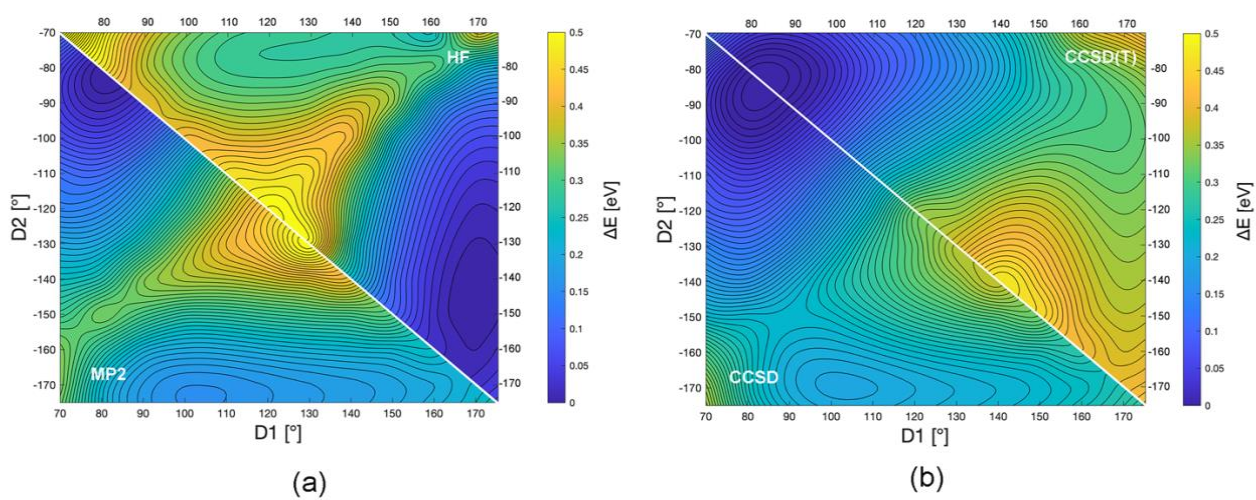

**Figure 3.** a) Energy surfaces calculated at the HF/aug-cc-pVDZ (upper) and MP2/aug-cc-pVDZ (lower) level of theory. b) The surfaces calculated at the CCSD/aug-cc-pVDZ (lower) and CCSD(T)/aug-cc-pVDZ (upper).

For the HF calculations and subsequently all post-HF calculations discussed, it was found to be crucial to check the stability of the converged HF-SCF solution via stability analysis (involves the computation of the electronic Hessian[82,83]). Thus, when the stability of the electronic state of each point on the HF surface was checked, it was found that all SCF solutions that indicated a delocalized electronic structure (approximately equal spin population on the nitrogen atoms) were actually found to be unstable solutions (one or more negative eigenvalues of the electronic Hessian). When the eigenvector associated with the negative eigenvalue was used to generate a new orbital guess, a lower energy solution was found, that had a localized electronic structure (spin density primarily on one nitrogen). The HF/aug-cc-pVDZ surface with all points corresponding to stable SCF solutions in Figure 3a thus shows a complete absence of a minimum corresponding to a delocalized state (in the $D1=-90°$, $D2=90°$ coordinate region). This result is in sharp contrast to previous theoretical results on $DMP_+$ that have highlighted the absence of the localized state instead for many DFT methods (compare e.g. to the B3LYP surface in Figure 2a). The HF surface shows the presence of the localized state DMP-L$_+$ (in the $D1=-145°$, $D2=170°$ coordinate region) and an additional conformer with a localized electronic structure, DMP-L2$_+$, can be seen in the $D1=-74°$, $D2=145°$ coordinate region. Electron correlation is hence not needed to describe the DMP-L$_+$ state since it already occurs on the HF surface;[84] in contrast to DMP-D$_+$ state, which instead on the HF surface features a high-energy ridge in the $D1=-90°$, $D2=90°$ coordinate region, that connects to the localized state involving the other nitrogen site. It appears that the symmetry of the system makes it easy for points on the diagonal line (equal D1 and D2 angles) to converge to the symmetric, delocalized, but ultimately unstable SCF solution at the HF level of theory. On the stable HF surface, the diagonal corresponds to the crossing of the two localized SCF solutions, that are equal in energy at those points. This HF instability encountered for the DMP-D$_+$ region has consequences for the use of post-HF methods for this molecule, as the use of an unstable

SCF determinant is a highly questionable reference determinant. It seems likely to us that recent previous HF and post-HF calculations,[13,14] of the DMP-D+ state have also involved an unstable SCF solution. We do note; however, that this HF instability for DMP+ was first reported on by Brouwer et al.[9] The MP2/aug-cc-pVDZ surface, also shown in Figure 3a, was calculated using stable HF (localized) solutions for every point. The surface shows two DMP+ minima separated by well-defined saddle point region. The Mulliken population analysis of the relaxed MP2 density showed that the point corresponding to the delocalized geometry (D1=-80°, D2=80° coordinates), has close, though not completely equal, spin populations on the nitrogen atoms, while the HF result indicates a clearly localized solution (see Table S5 in SI). Thus the MP2 wavefunction expansion appears able to correct for some (but not all) of the deficiency of the (localized) HF reference determinant in describing the delocalized state. The MP2 energy difference between DMP-D+ and DMP-L+ (using fully relaxed BHLYP structures) computed using the stable reference HF determinant is 0.153 eV (Table 1), only 0.033 eV smaller than the BHLYP energy. The occurrence of the unstable HF solution likely explains the large energy difference (0.82 eV) between DMP-D+ and DMP-L+ states previously reported on at the MP2 level of theory.[13]

*3.5.2 Coupled cluster theory surfaces*

Energy surfaces were calculated at the coupled cluster level of theory using either singles doubles expansion (CCSD) or the singles doubles perturbative triples expansion (CCSD(T)). All surface points were performed on top of stable (localized) HF solutions (as confirmed by a stability analysis on each point) and the surfaces are shown in Figure 3b. This CCSD surface is overall comparable to the MP2 surface (Figure 3a) and BHLYP (Figure 2a), although in the CCSD and MP2 surfaces the DMP-L+ minima are shifted to the lower D1 values in comparison to BHLYP. The two minima are directly visible on the surface, as is the saddle point region, which is consistent with the results of previous CCSD geometry optimizations locating two minima.[13,14] Spin population analysis of the unrelaxed CCSD density for the D1=-90, D2=90° point (see Table S5) confirms that similarly to MP2, the electronic structure is mostly delocalized with CCSD, despite being expanded from a HF reference wavefunction. Some symmetry-breaking remains with CCSD, though less than with MP2.

Thus, the CCSD wavefunction appears to be flexible enough to describe the electronic structure and energy surface of DMP+. The energy difference between DMP-D+ and DMP-L+ is calculated to be +0.23 eV[13] which is closer to the experimental value than BHLYP and MP2. However, remarkably when perturbative triples excitations are added, the potential energy surface changes dramatically. The CCSD(T) surface shows only one minimum (in clear disagreement with experiment), analogous to the results of most DFT methods and superficially appears to suffer from a delocalization problem analogous to B3LYP. This is an unusual result for CCSD(T), a generally highly accurate wavefunction theory method.

**3.4 Multi-reference wavefunction surfaces**

The unusual behaviour of HF theory for DMP+ (complete absence of the DMP-D+ state) and the inconsistent behaviour of post-HF methods (MP2, CCSD and CCSD(T)) suggests that the DMP+ cation may in fact be a true multireference system. The starting point for multireference wavefunction theory calculations is traditionally a CASSCF wavefunction and CASSCF calculations with different active spaces will be discussed first before discussing multireference dynamical correlation treatments (NEVPT2 and MRCI).

*3.4.1 CASSCF and DMRG-CASSCF calculations*

The active space is a critical choice for a reliable CASSCF calculation. As defined in the Computational details (and shown in the SI), multiple active spaces were calculated of different sizes and composition. A minimal active space for DMP+ contains only the nitrogen lonepairs (CASSCF(3,2)). As previously discussed, as most methods predict spin density between the C-C bonds, the C-C σ bonding/antibonding orbitals are arguably important as well and as the geometric comparison in Table S1 indicated, N–CH$_3$ σ or even N-C bridging σ bonding orbitals may also be required for a balanced treatment. To explore the convergence of the CASSCF wavefunction with active space size and composition we calculated energy surfaces using all defined active spaces shown for the CASSCF(7,8) and CASSCF(7,8)' in Figure 4a, the DMRG-CASSCF(19,20) in Figure 4b, while Figures S9a-b in the SI contain the surfaces for other active spaces.

Starting from the minimal CASSCF(3,2) wavefunction, one might expect this 2-configuration- state-function wavefunction to be flexible enough to describe the DMP-D+ region; however, its electronic structure was found to be symmetry-broken and highly localized according to Mulliken spin populations ( $s_{N1}$ =0.03, $s_{N2}$ = 0.79), even in the delocalized region. Furthermore, it was not possible to calculate the energy surface with a consistent active space for each point, suggesting this minimal CASSCF wavefunction to be too small.

Increasing the active space to CAS(7,8) resulted in more consistent energy surfaces for DMP+. The CASSCF(7,8) and the CASSCF(7,8)' surfaces are shown in Figure 4a where the active spaces differs (C-C σ bonding/antibonding orbitals in (7,8)' active space instead of N-C in (7,8)). Interestingly, the CASSCF(7,8) and CASSCF(7,8)' surfaces do not show DMP-D+ minima along the diagonal or along the (-90°, 90°) point, the minima instead being slightly shifted away from the diagonal. Analysis of the spin populations (Table S5) of the minimum energy (-100°, 80°) points indicated strong symmetry breaking and a localized electronic structure. Almost complete delocalization; however, was seen for the (90°, -90°) point for both CASSCF(7,8) and CASSCF(7,8)' (see Table S5).

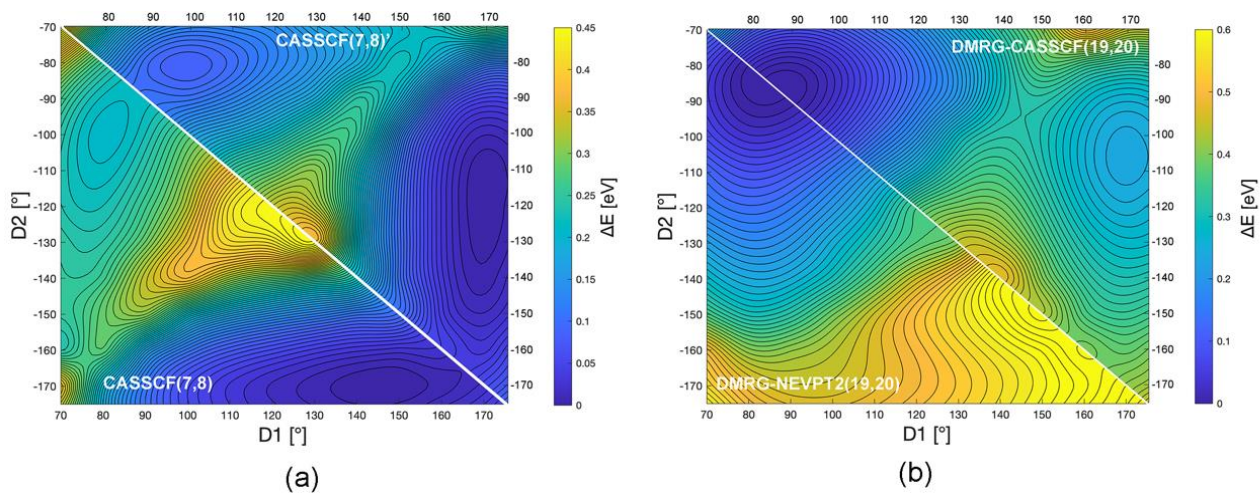

**Figure 4.** Energy surfaces calculated using (a) the CASSCF(7,8)'/aug-cc-pVDZ (upper) and CASSCF(7,8)/aug-cc-pVDZ (lower) and (b) the DMRG-CASSCF(19,20)/aug-cc-pVDZ (upper) and DMRG-NEVPT2(19,20)/aug-cc-pVDZ (lower) levels of theory calculated on the relaxed BHLYP structures.

Thus, to describe both the delocalized and localized regions of the energy surface well, we suspected it to be necessary to include both occupied and virtual orbitals associated with C-C bonds and N-CH$_3$ bonds in the active space. Indeed, the CASSCF(11,12)/aug-cc-pVDZ surface, shown in Figure S9b, appears to describe the delocalized region and localized regions well, giving an energy surface superficially similar to CCSD. However, the localized minimum is lower in energy than the delocalized minimum, giving a wrong energy difference (as seen in Table 1), suggesting the lack of some differential electron correlation still. To go beyond the CASSCF(11,12) active space and including the missing orbitals associated with the bridging N-C bonds, the CASSCF(19, 20) active space was investigated at the DMRG-CASSCF level. The energy surface was calculated at the DMRG-CASSCF(19,20)/aug-cc-pVDZ level shown in Figure 4b. The DMRG-CASSCF(19,20) surface reveals two well-defined minima and a clear saddle-point region, overall similar to the BHLYP surface and is consistent with experiment regarding stability of states.

The energy difference between the DMP-D$_+$ and DMP-L$_+$ structures relaxed at the DMRG-CASSCF(19,20)/aug-cc-pVDZ level of theory is 0.251 eV (Table 1). Importantly, the results are not very sensitive to the basis set size, with the energy difference changing to $\Delta E = 0.24$ eV at the aug-cc-pVTZ level. The saddle point connecting DMP-D$_+$ and DMP-L$_+$ was found to lie 0.07 eV above DMP-L+ at the DMRG-CASSC(19,20)/aug-cc-pVTZ level of theory.

*3.4.2 Multireference perturbation theory and multireference configuration interaction*

Multireference perturbation theory is the most straightforward way of accounting for dynamic correlation from a reference CASSCF wavefunction, with CASPT2 and NEVPT2 being the most common methods. NEVPT2 calculations were performed on top of the CASSCF reference wavefunctions for all active spaces and surfaces previously described. The NEVPT2 results were found to be strongly dependent on the size of the active space and when NEVPT2 calculations on CASSCF(7,8), CASSCF(7,8)' reference wavefunctions were used, the missing DMP-D$_+$ minimum at the CASSCF level, reappeared (Figure S10a, S10b). This suggests that the missing electron correlation effects at the CASSCF level with these smaller active spaces are reasonably well introduced via the NEVPT2 approach; however, the energy difference between DMP-D$_+$ and DMP-L$_+$ was found to be strongly overestimated (DMP-D$_+$ too stable) as seen in Table 1. When NEVPT2 calculations were performed on top of CASSCF reference wavefunctions with even larger active spaces there were more surprises. The DMP-L$_+$ minimum is barely visible and very shallow when using CASSCF(11,12) reference wavefunction as seen in Figure 2b. When NEVPT2 calculations are carried out on the DMRG-CASSCF(19,20) wavefunction the energy difference is in fair agreement with experiment (0.45 eV); however, the DMP-L+ minimum is not visible on the DMRG-NEVPT2(19,20) surface (Figure 4a).

As going beyond perturbation theory is clearly necessary, we performed calculations at the multireference configuration interaction (MRCI+Q) level, using a fully internally contracted MRCI approach as implemented in ORCA (FIC-MRCI+Q) using the same CASSCF reference wavefunctions previously described. The cc-pVDZ basis set was used for all FIC-MRCI+Q PES surfaces. The energy surfaces calculated at the FIC-MRCI+Q level for small active spaces ((3,2), (7,8), (7,8)') are shown in Figure S10a-b in the SI and compared to the NEVPT2 surfaces with the same reference wavefunctions. While the shapes of the MRCI+Q and NEVPT2 surfaces are qualitatively similar, the energy differences between delocalized and localized wells are very different with the NEVPT2 energy differences larger compared to the FIC-MRCI+Q results (Table 1). Reassuringly, as previously mentioned, when FIC-MRCI+Q calculations are performed for the energy surface using a CASSCF(11,12) reference wavefunction, the surface, shown in Figure 2b, reveals clear minima for both the delocalized and localized regions, in sharp contrast to the behaviour of the NEVPT2-CASSCF(11,12) surface. This suggests that the problems associated with the disappearance of minima at the NEVTP2 levels for CASSCF(11,12) and DMRG-CASSCF(19,20) are associated with the perturbation treatment employed. FIC-MRCI+Q calculations on top of a DMRG-CASSCF(19,20) reference wavefunction would be a desirable future goal but based on the trends seen for the these surfaces, we do not expect the FIC-MRCI+Q results to fundamentally change with an even larger active space. As seen in Table 1, the energy differences between DMP-D+ and DMP-L+ minima at the FIC-MRCI+Q level (using BHLYP structures) are much less sensitive to the size of the active space than the NEVPT2 results and show that a level of convergence (~0.03 eV) obtained with CASSCF(7,8)'/CASSCF(11,12) references. The largest calculation performed, FIC-MRCI+Q using a CASSCF(11,12) reference wavefunction with the aug-cc-pVDZ basis set on the BHLYP relaxed DMP+ minima gives an energy difference of 0.336 eV which is remarkably close to the experimental value of 0.33 eV. While these FIC-MRCI+Q/aug-cc-pVDZ calculations are probably not completely converged with respect to basis set size, NEVPT2 calculations with the aug-cc-pVDZ and aug-cc-pVTZ basis sets with several different active spaces suggest basis set effects on the energy difference to be quite systematic and on the order of 0.04-0.08 eV at most.

## 4. DISCUSSION

The DMP cation is an unusual molecule where the potential energy surface dramatically changes depending on the theory level. While problems linked to delocalization error in DFT approaches have been described for other mixed-valence cations, DMP+ is an unusually difficult case. In a previous study,[13] density functional theory methods were found to fail to describe the potential energy surface of DMP+ apparently due to overdelocalization, and presumably this is related to the general problem of self-interaction and delocalization error in density functional theory. The half-and-half (50 % HF exchange) hybrid BHLYP functional performs well, seemingly due to being able to balance the overdelocalized description by DFT and the overlocalized description of HF and shows both minima on the surface as shown in Figure 2a, in excellent agreement with MRCI+Q calculations in Figure 2b. However, as shown in the Results section, DMP+ is not only a case where density functional theory approximations struggle, but emerges as a truly challenging case for correlated wavefunction theory as well.

The comparison between energy surfaces calculated at the CASSCF level with different active spaces gives important insight into how electron correlation affects the electronic structure of this mixed-valence cation. A poor or imbalanced electron correlation treatment breaks the symmetry of the wavefunction in the delocalized structure region, which has consequences not only in giving a wrong electronic structure, but results in artificial localized minima instead, visible on the HF, CASSCF(7,8) and CASSCF(7,8)' surfaces. Both C-C and N-CH$_3$ bonding and antibonding orbitals appear to be crucial for a balanced description of both DMP$_+$ states (the bridging C-N bond orbitals being somewhat less important), with the large CASSCF(19,20) active space, that includes all orbitals associated with C-C and C-N bonds giving the most balanced treatment according to the energy surface and the energy difference between the two states. These large active-space CASSCF calculations appear to give consistent results that compare well with experiment and are converged with respect to basis set. However, CASSCF wavefunctions primarily describe the static correlation present in the system. While expanding the active space to CASSCF(19,20) captures some of the dynamic correlation (possibly most of the differential dynamic correlation between the two states), approximately 1 Hartree (27 eV) of dynamic correlation energy is still not accounted for (based on NEVPT2 and MRCI results). Thus, the DMRG-CASSCF calculations can only be giving the correct result if the remaining correlation energy for both DMP-D$_+$ and DMP-L$_+$ species is approximately equal and cancels out. We note in this context that there has been considerable success recently in using large-active space DMRG calculations for e.g. spin coupling problems without including dynamic electron correlation,[85] relying on similar systematic error cancellation.

The troubling NEVPT2 results especially for the (19,20) active space suggest severe problems with multireference perturbation theory for describing mixed-valence cations such as DMP$_+$ and upon consulting the literature, we found similar documented problems with other systems. Overall, the multireference wavefunction theory results for DMP$_+$ are very sensitive to the nature of the reference wavefunction and how the dynamic correlation is described. The fact that the NEVPT2 method, using the largest possible CASSCF reference wavefunction (that gives a well-behaved surface on its own), results in an energy surface of DMP+ that is inconsistent with experiment. Using a very large active space is troubling; however, and suggests a general failure of using perturbation theory in describing the potential energy surface of mixed-valence systems such as DMP+. These problems are likely related to similar problems discussed in the literature for the "Spiro" mixed-valence cation, which was extensively investigated using multiconfigurational and multireference methods.[41,42,43,44,45] Helal et al.[41] studied the "Spiro" cation via CASSCF calculations and multireference CI and described it as a Robin-Day Class II localized system. Severe problems with CASPT2 and NEVPT2 were found[42,43] as the ground state exhibited an unphysical symmetrical well at the CASPT2/NEVPT2 level where a double-well was expected. While third-order perturbation theory (NEVPT3) resolved the problem this was prohibitively expensive.[43]

Turning to the behaviour of the single-reference methods, the absence of the DMP-D+ state on the HF potential energy surface clearly has consequences for a post-HF wavefunction treatment. As the DMP-D+ state cannot be described via a regular single HF determinant, this calls for a multireference treatment and the DMP-D+ state is thus arguably by definition a multireference problem. Single-reference wavefunction theory operates on the principle that the HF wavefunction is a suitable reference wavefunction for the system. This is clearly not the case for DMP+. Previous theoretical studies[13,14] have presented MP2, CCSD and CCSD(T) calculations of DMP+. In view of the unstable HF solutions found in our work, it seems likely that previous calculations of DMP-D+ have, without knowledge, been performed on unstable HF solutions, and the reliability or usefulness of such a calculation is not clear. It is also not clear from the outset whether a stable HF determinant wavefunction of a localized state is a useful reference determinant at all for the wavefunction expansion of the delocalized state. In this context it is worth noting recent attempts to improve HF theory via removing complex-conjugation of orbital coefficients resulting in Holomorphic Hartree-Fock methods.[86,87] This approach has been used successfully to describe charge transfer in similar donor-acceptor systems.[88] While our previously discussed MP2 results (Figure 3a) do show a qualitatively correct energy surface, the electronic structure is not completely symmetrical according to a population analysis of the MP2 relaxed density (see Table S5 in SI).

In a recent comment[14], Ali *et al.* discuss CCSD and CCSD(T) calculations of DMP+ and compare DFT results to these calculations, using the CCSD(T) results as a high-level reference theory, and argue that DMP+ does not represent a failure of common density functional approximations. As is now clear, this is a problematic comparison as there does not appear to be a minimum for the DMP-L+ state on the CCSD(T) potential energy surface (Figure 3b) and hence there is no well-defined energy difference between minima to speak of. The lack of a stable delocalized solution at the reference HF level and the contrasting behaviour of CCSD and CCSD(T) clearly makes single-reference coupled cluster theory not suitable as a high-level reference theory. In order to understand whether the behaviour of CCSD(T) stems from the flawed UHF reference function or perhaps the perturbative triples correction, we performed CCSD(T) calculations using alternative reference wavefunctions, as detailed in the SI. CCSD(T) calculations using Brueckner orbitals,[89] quasi-restricted orbitals[90] or UKS-DFT orbitals did not, however, lead to an improved CCSD(T) surface; see SI for a discussion. We also performed orbital-optimized coupled cluster theory (OO-CCD(T)) calculations, where the orbitals are variationally optimized at the CCSD level (instead of the HF level), thus effectively removing any effect of the HF reference. While this encouragingly gave a relaxed OO-CCD density that showed complete delocalization (no symmetry-breaking) according to spin population analysis of the D1=-90, D2=90° point (See Table S5) , the OO-CCD(T) surface is still missing the DMP-L+ minimum (like UHF-CCSD(T)) as shown in the Figure S11e in SI. These results suggest that the HF reference wavefunction is not the main problem in the CCSD(T) wavefunction, but rather the wavefunction expansion itself, perhaps due to an imbalance in the static and dynamic correlation of the wavefunction when the perturbative triples correction is added. It would be of much interest to see whether these problems are resolved at the (very expensive) CCSDT level (full triples)[91] or CCSDT(Q)[92] levels (full triples and perturbative quadruples) or with alternative triples approximations.[93,94,95,96,97,98] We note that this inaccurate behaviour of CCSD(T) is in contrast to the study of the problematic mixed-valence system [Al$_2$O$_4$]- by Kaupp *et al.*[22] where perturbative triples excitations made a large contribution to the correlation energy, but CCSD(T) was overall well-behaved compared to CCSDT(Q). In view of the previously discussed problems with multireference perturbation theory we suggest the perturbation treatment of the triples excitations as the reason for the failure of CCSD(T).

## 5. CONCLUSIONS

We have presented an extensive theoretical investigation of the DMP cation and shown that the molecule represents an unusual challenge to black-box quantum chemistry, whether at the density functional theory level or the wavefunction theory level.

While a systematic expansion of the active space in CASSCF calculations leads ultimately to an energy surface in agreement with experiment, the inclusion of dynamic correlation via NEVPT2 introduce active-space dependent artefacts, which, however, are absent in the more robust multireference configuration interaction treatment. Both large-active space CAS(19,20) DMRG-CASSCF calculations and smaller active-space CAS(11,12) FIC-MRCI+Q calculations give a consistent picture of the potential energy surface and are in good agreement with the experimental estimate for the reaction energy and produce a significant reaction barrier consistent with the measurements.

Remarkably, while CCSD calculations are qualitatively correct, CCSD(T) calculations are not, suggesting a rare failure of the perturbative triples correction.

The DMP cation, despite its apparent simplicity, is hence a truly challenging case for correlated wavefunction theory and the presented results should make the system useful as a benchmark system for the study of electronic state localization and could guide the development of more robust and affordable correlated wavefunctions methods as well as density functionals.

**Acknowledgements**

MG acknowledges post-doctoral fellowship from the University of Iceland Research Fund. This work was supported by the Icelandic Research Fund. The computations were performed on resources provided by the Icelandic High Performance Computing Centre at the University of Iceland.

# Supporting Information

# Localized and delocalized states of a diamine cation: A critical test of wave function methods

*Marta Gałyńska[a], Vilhjálmur Ásgeirsson[a], Hannes Jónsson[a]\*, Ragnar Bjornsson[b]\**

[a] Faculty of Physical Science, VR-III, University of Iceland, 107 Reykjavik, Iceland

[b] Max-Planck Institute for Chemical Energy Conversion, Stiftstrasse 34-36, 45470 Mülheim an der Ruhr, Germany.

**\*Corresponding author**: ragnar.bjornsson@cec.mpg.de, hj@hi.is

Table of Contents

# 1. Molecular and electronic structures of neutral DMP and its cation.

**Table S1.** The C-C bond length average [Å], the D1 and D2 angles values [°] of neutral DMP, DMP-D+ and DMP-L+ relaxed at different levels of theory using the aug-cc-pVDZ basis set.

|         | Method              | C-C [Å] | D1 [°]  | D2 [°] |
|---------|---------------------|---------|---------|--------|
| DMP-1   | BHLYP               | 1.52    | -175.2  | 175.2  |
| DMP-2   | BHLYP               | 1.52    | -171.9  | 86.8   |
| DMP-3   | BHLYP               | 1.53    | -75.1   | 75.0   |
| DMP-L+  | HF                  | 1.53    | -145.1  | 169.5  |
|         | BHLYP               | 1.53    | -132.2  | 169.5  |
|         | CASSCF(7,8)         | 1.52    | -148.8  | 169.8  |
|         | CASSCF(7,8)'        | 1.56    | -120.5  | 169.9  |
|         | CASSCF(11,12)       | 1.56    | -116.0  | 170.5  |
|         | DMRG-CASSCF(19,20)  | 1.57    | -104.5  | 169.5  |
|         | CCSD[a]             | 1.56    | -103.3  | 170.2  |
| DMP-D+  | HF (unstable)[a]    | 1.62    | -88.7   | 88.7   |
|         | B3LYP               | 1.61    | -90.2   | 90.2   |
|         | BHLYP               | 1.60    | -88.5   | 88.5   |
|         | CASSCF(11,12)       | 1.65    | -87.9   | 87.9   |
|         | DMRG-CASSCF(19,20)  | 1.65    | -87.3   | 87.3   |
|         | CCSD[b]             | 1.62    | -84.6   | 84.6   |
| DMP-SP+ | BHLYP               | 1.56    | -97.3   | 151.8  |
|         | DMRG-CASSCF(19,20)  | 1.61    | -92.1   | 145.9  |

[a] As discussed in the manuscript, the DMP-D+ state is an unstable SCF wavefunction at the HF level and there is not a stable minimum for a delocalized state on the HF surface. The shown geometry corresponds to the relaxed structure of this unstable wavefunction.
[b] The relaxed structures were taken from Ref. 13.

The three conformers of neutral DMP molecules relaxed at the BHLYP/aug-cc-pVDZ level of theory are shown in Figure S1. The chair structure of neutral DMP (DMP-1) as calculated at the BHLYP/aug-cc-pVDZ level of theory has both D1 and D2 angles close to 180° as both methyl groups are in the equatorial position; this conformer would be the majority conformer of neutral DMP under gas phase experimental conditions.[84,85,86] Two other chair conformers of neutral DMP exist, with one axial methyl group, DMP-2 (0.242 eV), or both methyl groups in the axial positions, DMP-3, (0.364 eV). There are also 3 twist conformers (>0.48 eV).

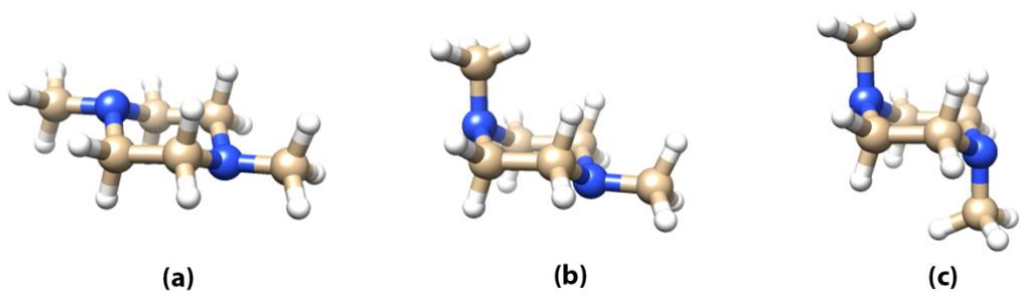

**Figure S1.** The three neutral conformers of DMP molecule: (a) DMP-1, (b) DMP-2, and (c) DMP-3 relaxed at the BHLYP/aug-cc-pVDZ level of theory.

The relaxed DMP-L$_+$ geometries with different methods showed some variation in the C-C bond lengths of DMP-L$_+$, but primarily in the D1 dihedral angles. The HF, BHLYP, CAS(7,8) structures had a rather wide D1 angle, ranging between -148.8° in CASSCF(7,8) to -132.2° in BHLYP. Changing the nature of the active orbitals, in going from CASSCF(7,8) to CASSCF(7,8)' calculations brought a significant reduction in the D1 value for 28.3° and a 0.03 Å increase in the C-C bond length. These geometrical changes are without doubt related to the presence of C-C σ and σ* orbitals in the CASSCF(7,8)' active space. Increasing the active space further to CASSCF(11,12), which includes both σ/ σ* orbitals used in the CASSCF(7,8) and CASSCF(7,8)' calculations, the D1 angle changes only by 0.5° with respect to the CASSCF(7,8)' structure and marginal changes in the C-C bond lengths. Using the DMRG-CASSCF approach it was possible to explore a large CASSCF(19,20) active space which brought further changes to the D1 value, decreasing by 15.4° compared to CASSCF(11, 12). Interestingly there is very good agreement between DMRG-CASSCF(19,20) and CCSD structures (dihedral angle difference of 1.2°), despite the completely different nature of the correlation treatment. Finally we note that CASSCF(11,12) and DMRG-CASSCF(11,12) result in identical structural parameters, demonstrating the approximate DMRG-CASSCF wavefunction as well as the approximate analytical gradient used in the DMRG-CASSCF calculations is accurate (see Table S1). These results reveal the importance of electron correlation in describing the electronic structure of the DMP-L$_+$ state (as shown via the geometry). A CASSCF(7,8) active space results in only small structural changes compared to HF (thus not capturing the necessary correlation), while the CASSCF(7,8) and CASSCF(7,8)' geometric comparison (and compared to larger active spaces) reveals the importance of the C-C σ/ σ* orbitals in describing the correlation of this state. These geometrical changes could be correlated to the changing nature of the natural orbitals (orbitals obtained via diagonalization of the density matrix) associated with the non-bonding nitrogen orbitals as a function of correlation treatment. This comparison, shown in Figure S2, reveals how the nitrogen orbitals show rather localized character at the HF level (some mixing with H orbitals) with only small changes brought in by the CASSCF(7,8) correlation treatment. Changing the active space CASSCF(7,8)' (bringing in C-C bond orbitals), however, results in the mixing-in of C-C character into the ionized lonepair (leftmost N site on the figure) and this is then reflected in the large change in the D1 angle with respect to the CASSCF(7,8) calculation. Further increasing the active space by addition of C-N σ/σ* orbitals in the CASSCF(11,12)/DMRG-CASSCF(19,20) calculations results in mixing of the unionized N lone pair with σ C-N as well as H$_3$C-N orbitals, which brings further changes to the DMP-L$_+$ geometry. Interestingly, the natural orbitals from a relaxed MP2 density and an orbital-optimized coupled cluster density (OO-CD), the unionized lone pair looks very similar to the HF and BHLYP orbitals, while the ionized lone pair mixes with σ C-C as well as H$_3$C-N orbitals.

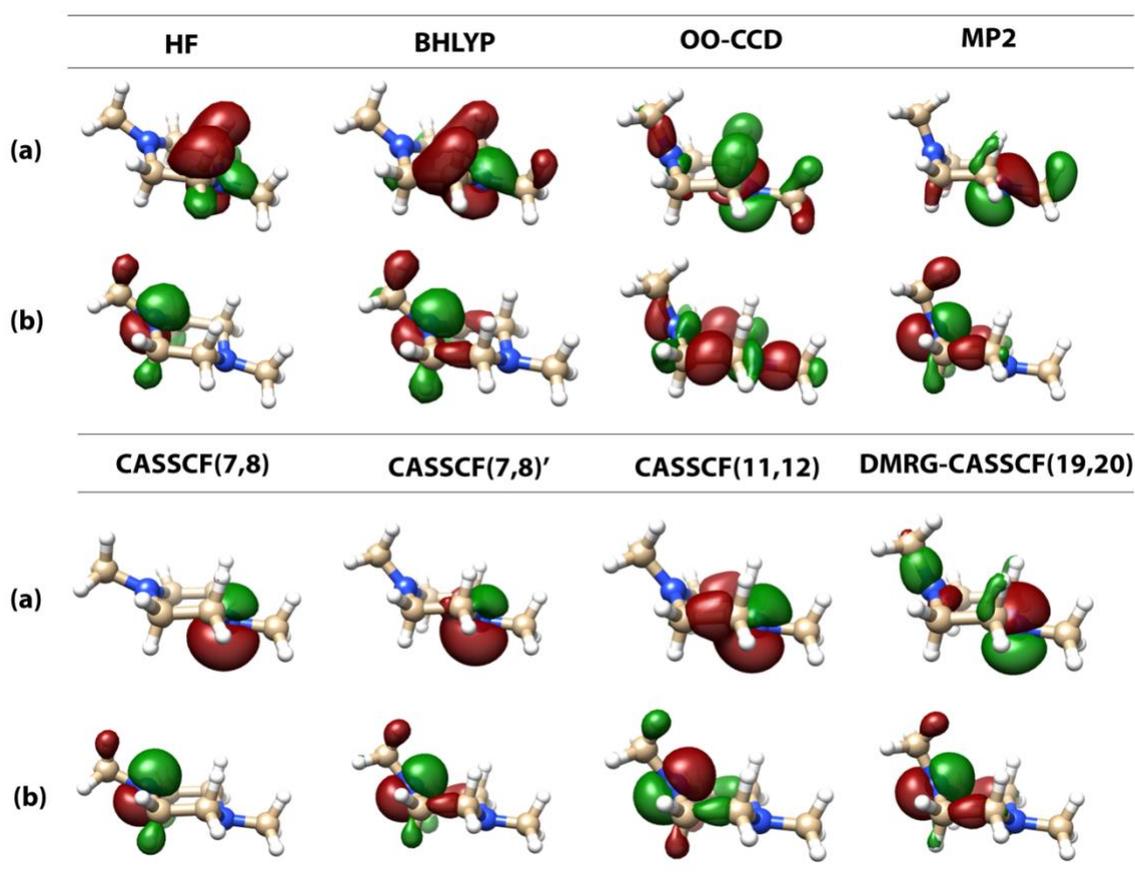

**Figure S2.** The natural orbitals associated with the nonbonding nitrogen lonepairs calculated for the relaxed DMP-L+ structures (at each level) using the HF, BHLYP, OO-CCD, MP2 (using a relaxed density), CASSCF(7,8), CASSCF(7,8)', CASSCF(11,12), and DMRG-CASSCF(19,20) methods with the aug-cc-pVDZ basis set. The orbitals were plotted using isosurface level equal to 0.05. The orbitals in row (b) are the site of the hole (according to the local nitrogen geometry).

In the DMP-D+ structure, the delocalization of the unpaired electron brings further geometrical changes. These structural changes reflect the delocalized electronic structure and the more sp2-like character of both nitrogen atoms. The DMP-D+ structure has $C_{2h}$ symmetry and is an unusual axial-axial conformer, closely resembling the neutral DMP-3 conformer (see Figure S1) but showing a noticeable pseudopyramidal geometry (less sp3 character) of each nitrogen site and noticeably longer C-C bond lengths (0.07-0.09 Å longer at the BHLYP/aug-cc-pVDZ level) than the neutral DMP conformers. The presence of DMP-D+ will naturally depend on the electronic coupling strength of the two redox sites (lonepairs). As previously discussed by Cheng et al.[53] the spin density is delocalized over both nitrogen atoms in DMP-D+ and also shows spin density between the carbon atoms in the ring as shown in Figure 1c. Interestingly spin density plots generated with different methods looked overall quite similar as shown in Figure S3.

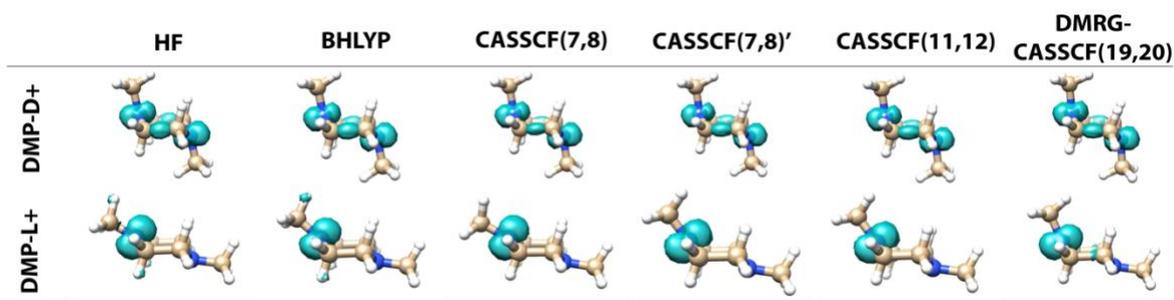

**Figure S3.** The spin density isosurfaces calculated at HF, BHLYP, CASSCF(7,8), CASSCF(7,8)', CASSCF(11,12), and DMRG-CASSCF(19,20) levels of theory. The density plots used isosurface levels of 0.01 e/Å$^3$.

The DMP-D$_+$ structures calculated with various methods as shown in Table S1 are overall quite similar with respect to the dihedral angles, differing most by 4.8°. However, significant differences can be seen for the C-C bond lengths. The CCSD structure by Cheng et al.[53] features a C-C bond length that is 0.05 Å longer compared to the BHLYP structure. The B3LYP-calculated structure is similar to the BHLYP structure. As discussed in the manuscript, the DMP-D$_+$ state presents challenges to HF and CASSCF wavefunctions and a minimum featuring both a symmetric $C_{2h}$ structure and a delocalized electronic structure could not be located at the HF, CASSCF(7,8)', or CASSCF(7,8) level of theory. The CASSCF(11,12), DMRG-CASSCF(19,20) structures have C-C bond lengths 0.04-0.05 Å longer compared to the BHLYP/B3LYP structures with the CCSD structure being in between BHLYP and the largest CASSCF calculation. Clearly electron correlation has a significant effect on the geometry, particularly regarding the nature of the interaction between the N lonepairs, though it is not clear which electronic structure method is describing the DMP-D$_+$ state best.

The structure of the saddle-point DMP-SP$_+$ connecting DMP-D$_+$ and DMP-L$_+$ states is shown in Figure 1d. It was located using an eigenvector-following method at the BHLYP and DMRG-CASSCF(19,20) levels of theory. Both of those structures had one imaginary mode, -186.06 cm$_{-1}$ (BHLYP) and -196.28 cm$_{-1}$ (DMRG-CASSCF(19,20)), confirming it as a 1$_{st}$ order saddle point on the DMP$_+$ surface. Although the BHLYP and DMRG-CASSCF(19,20) DMP-L+ geometries displayed some significant differences in the D1 and D2 values, the DMP-SP$_+$ structures were very similar with respect to the dihedral angles, which varied only for 5.2° and 5.9° for the D1 and D2 angles, respectively. Similar to DMP-D$_+$ and DMP-L$_+$, the C-C bond values are shorter for 0.05 Å in the BHLYP DMP-SP$_+$ structure in comparison to DMRG-CASSCF(19,20) geometry.

## 2. Active spaces.

The active spaces used in CASSCF/NEVPT2/MRCI calculations are shown in Figures S4-S8.

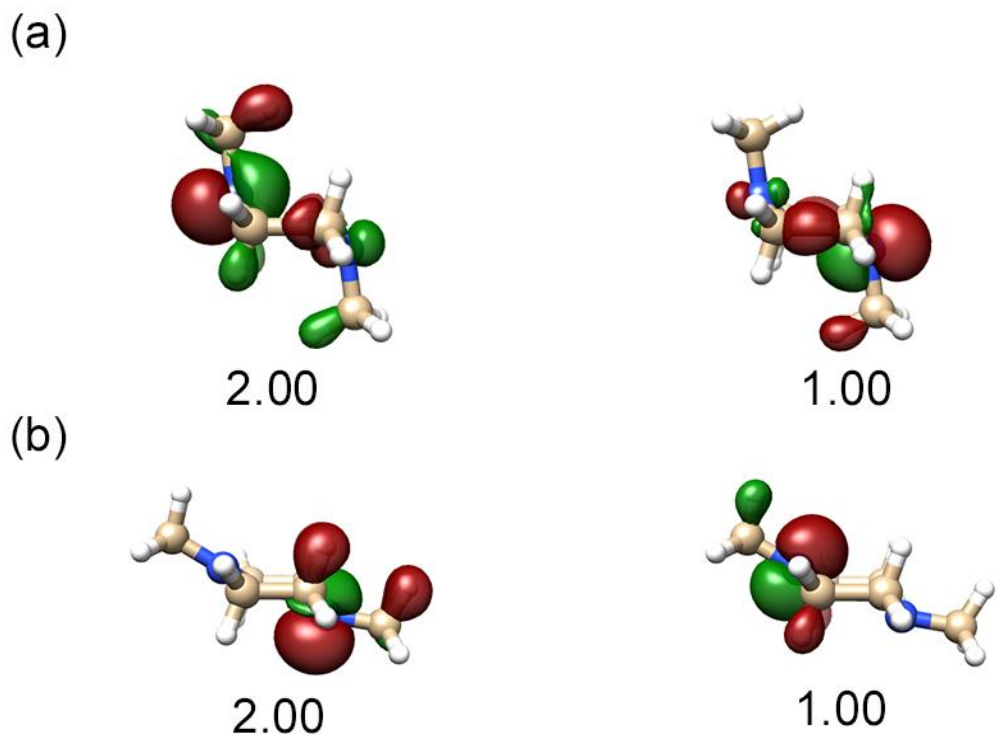

**Figure S4.** CAS(3,2) active space for DMP-D+ (a) and DMP-L+ (b). Shown are natural orbitals calculated at the CASSCF/aug-cc-pVDZ level of theory and their natural occupation numbers. This active space consists of the two N lone pair orbitals.

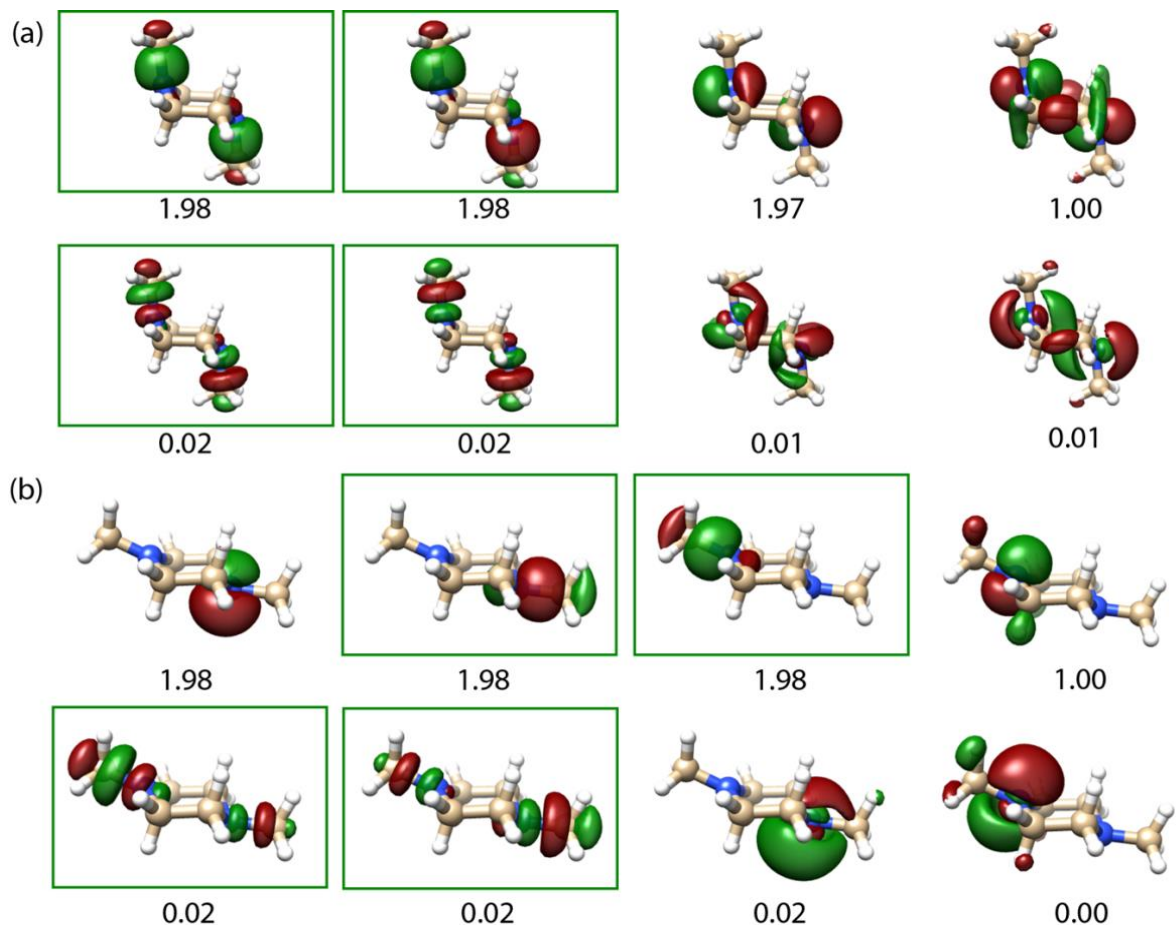

**Figure S5.** CAS(7,8) active space for DMP-D+ (a) and DMP-L+ (b). Shown are natural orbitals calculated at the CASSCF/aug-cc-pVDZ level of theory and their natural occupation numbers. This active space contains two N lone pairs and two σ H3C-N bonding orbitals and corresponding virtual orbitals. The σ H3C-N bonding and virtual orbitals are indicated in the green boxes.

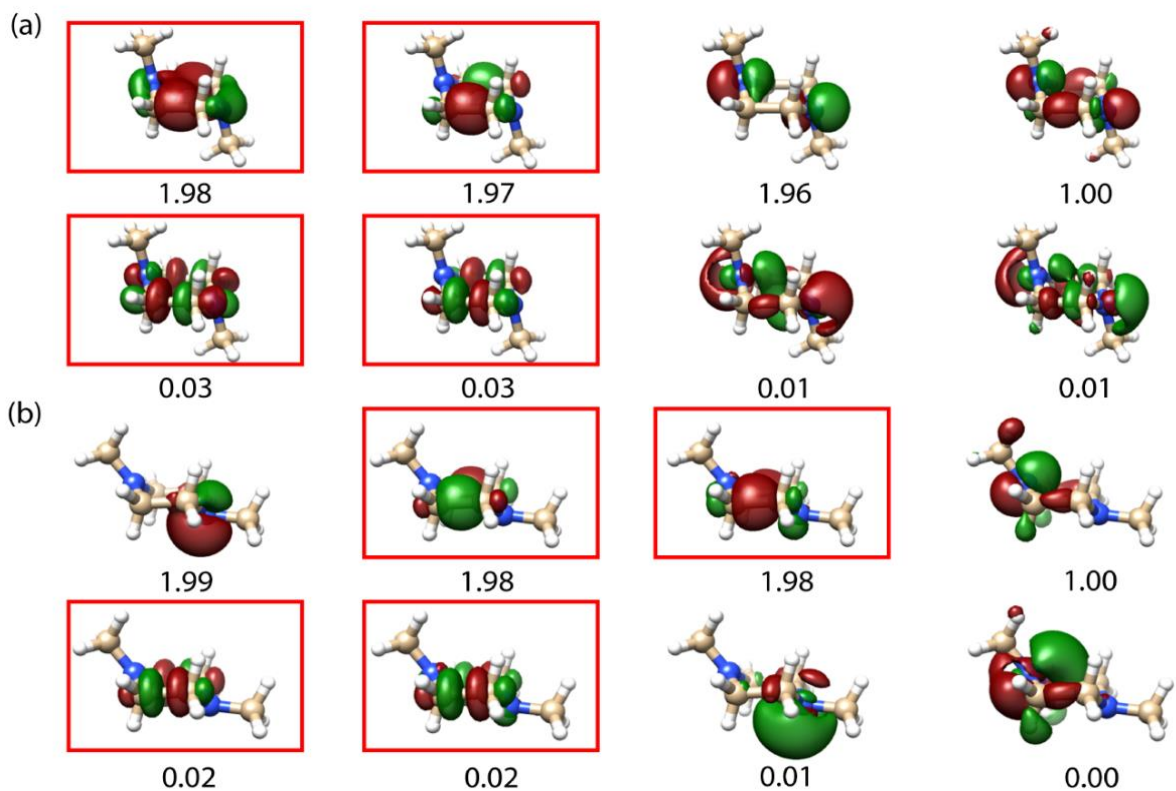

**Figure S6.** CAS(7,8)' active space for DMP-D$_+$ (a) and DMP-L$_+$ (b). Shown are natural orbitals calculated at the CASSCF/aug-cc-pVDZ level of theory and their natural occupation numbers. This active space consists of two N lone pairs, two σ C-C bonding orbitals, as well as, corresponding virtual orbitals. The σ C-C bonding and virtual orbitals are indicated in the red boxes.

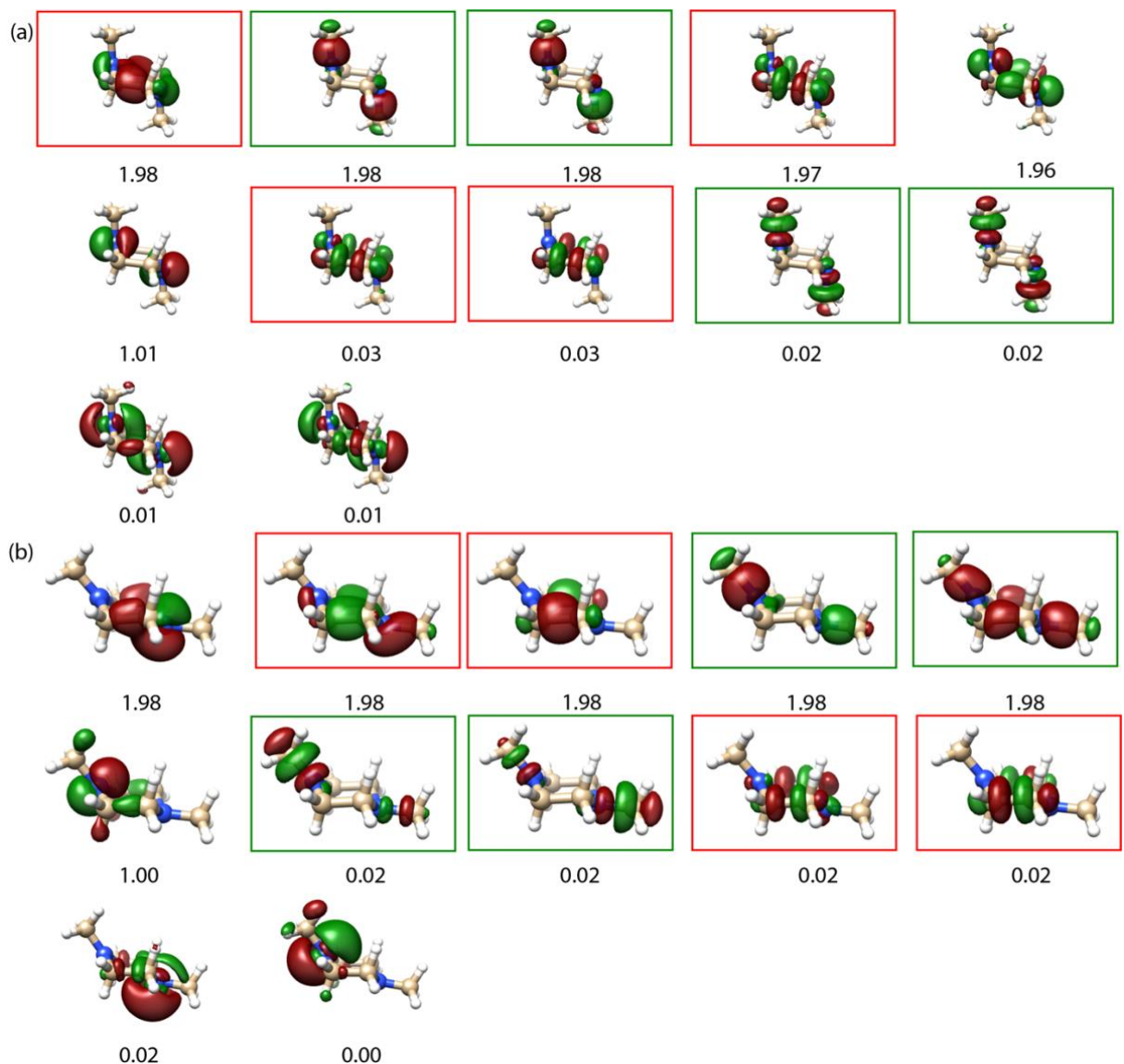

**Figure S7.** CAS(11,12) active space for DMP-D+ (a) and DMP-L+ (b). Shown are natural orbitals calculated at the CASSCF/aug-cc-pVDZ level of theory and their natural occupation numbers. The active space contains two N lone pairs, two σ H₃C-N, two σ C-C bonding orbitals and corresponding virtual orbitals. The σ H3C-N and C-C bonding and virtual orbitals are indicated in the green and red boxes, respectively.

(a)

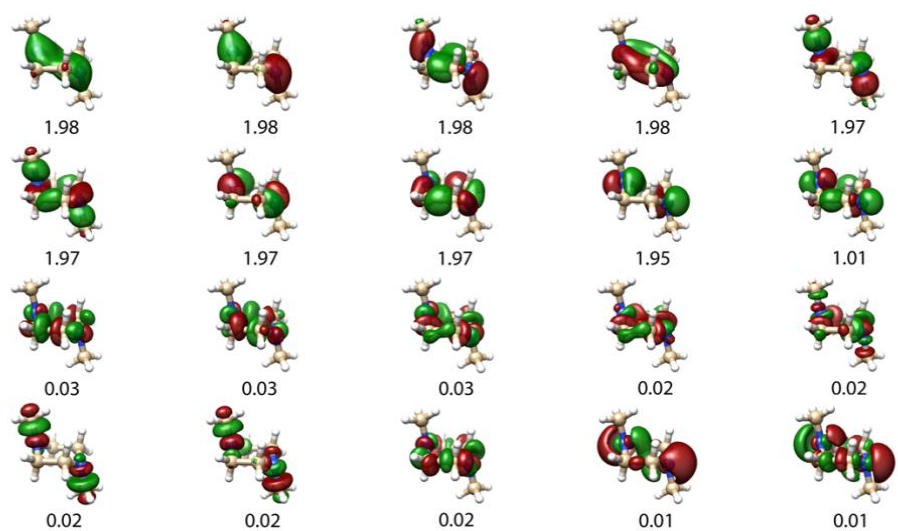

(b)

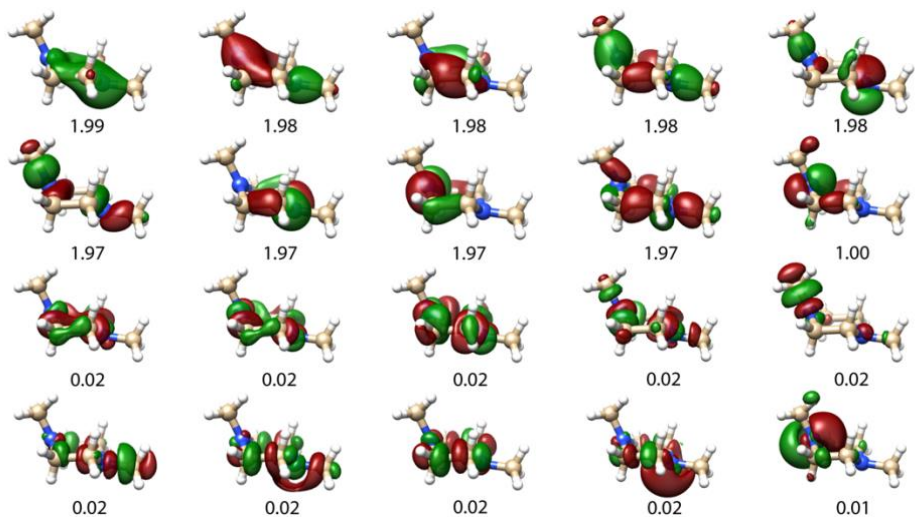

**Figure S8.** CAS(19,20)' active space for DMP-D+ (a) and DMP-L+ (b). Shown are natural orbitals calculated at the CASSCF/cc-pVDZ level of theory and their natural occupation numbers. The active space contains two N lone pairs, two σ $H_3C$-N, two σ C-C, four σ C-N bonding orbitals and corresponding virtual orbitals.

## 3. Total energies of DMP-D+ and DMP-L+.

Total electronic energies associated with the energy differences in Table 1 of main articles.

**Table S2.** Total energies [$E_h$] of DMP-D+, DMP-L+ , DMP-SP+ calculated at the different levels of theory.

| Method | DMP-D+ | DMP-L+ | DMP-SP+ |
| --- | --- | --- | --- |
| *Relaxed structures at each theory level* | | | |
| BHLYP/aug-cc-pVDZ | -346.103680383746 | -346.09710267699 | -346.0959004620 |
| DMRG-CASSCF(19,20)/aug-cc-pVDZ | -344.288198097204 | -344.27893064740 | -344.2763204713 |
| DMRG-CASSCF(19,20)/aug-cc-pVTZ | -344.37182679549 | -344.36313890693 | -344.3605101037 |
| *Single point FIC-MRCI+Q(11, 12)/aug-cc-pVDZ calculations on different geometries* | | | |
| CCSD/aug-cc-pVDZ | -345.242981034102 | -345.231161369666 | |
| DMRG-CASSCF(19,20)/aug-cc-pVDZ | -345.240538081157 | -345.228760909202 | |
| *Single point calculations on the relaxed BHLYP minima* | | | |
| MP2 | -345.2356301779 | -345.22997100773 | -345.226977397971 |
| CCSD | -345.3342028737 | -345.32604929059 | -345.325589011300 |
| CASSCF(11,12) | -344.1289828950 | -344.12961914000 | |
| DMRG-CASSCF(19,20) | -344.2831127557 | -344.27333195613 | |
| NEVPT2(3,2) | -345.2341980747 | -345.22917323070 | -345.225982805481 |
| NEVPT2(7,8) | -345.2368036440 | -345.20940125000 | |
| NEVPT2(7,8)' | -345.2519413608 | -345.21878474053 | |
| NEVPT2(11,12) | -345.2330799596 | -345.20474291756 | |
| MRCI(3,2) | -345.2049394121 | -345.21098343922 | |
| MRCI(3,4) | -345.2227495779 | -345.21546309125 | |
| MRCI(7,8) | -345.2333739345 | -345.22374422962 | |
| MRCI(7,8)' | -345.2310302037 | -345.21972829333 | |
| MRCI(11,12) | -345.2419100914 | -345.22955980326 | |

### 4. DMRG-CASSCF convergence tests.

Convergence of the DMRG-CASSCF wavefunction was tested with respect to the M parameter (number for renormalized states) for the (11,12) active space as presented in Table S3 and compared to the unapproximated CASSCF results. The tests showed that neither total energies or relative energies between DMP-D+ and DMP-L+ states are very sensitive with respect to the M parameter value in the range of M=500-1000 and are practically identical to the unapproximated CASSCF energies.

**Table S3.** Total energies [$E_h$] and relative energy differences [eV] between DMP-D+ and DMP-L+ states calculated at DMRG-CASSCF(11,12)/aug-cc-pVDZ level of theory for different M parameter values. Also shown are unapproximated CASSCF(11,12)/aug-cc-pVDZ results

| M<br>DMP state | 500 | 750 | 1000 | CASSCF(11,12) |
|---|---|---|---|---|
| DMP-D+ | -344.132501943771 | -344.132501965886 | -344.132501967398 | -344.132501968016 |
| DMP-L+ | -344.132776352224 | -344.132776356837 | -344.132776357055 | -344.132780494958 |
|  | -0.00744 | -0.00744 | -0.00744 | -0.00755 |

Because the DMRG-CASSCF analytical gradient as implemented in in ORCA is approximate (not completely obeying the Hellmann-Feynman theorem), the DMP+ structures obtained with DMRG-CASSCF(11,12) were compared with unapproximated CASSCF(11,12) structures. The comparison of the CASSCF and DMRG-CASSCF structural comparison in Table S4 showed that the structures are practically identical for the C-C bond length as well as D1 and D2 dihedral angles.

**Table S4.** The C-C bond length average [Å], the D1 and D2 angles values [°] of the DMP-D+ and DMP-L+ structures relaxed at CASSCF(11,12) and DMRG(11,12) levels of theory using the aug-cc-pVDZ basis set.

|  | Method | C-C [Å] | D1 [°] | D2 [°] |
|---|---|---|---|---|
| DMP-D+ | CASSCF(11,12) | 1.654697 | -87.927062 | 87.927173 |
|  | DMRG-CASSCF(11,12) | 1.654691 | -87.927585 | 87.927705 |
| DMP-L+ | CASSCF(11,12) | 1.564937 | -116.036676 | 170.498965 |
|  | DMRG-CASSCF(11,12) | 1.564934 | -116.036670 | 170.497996 |

## 5. Energy surfaces with additional active spaces at CASSCF/NEVPT2/MRCI levels of theory.

The CASSCF/NEVPT2/FIC-MRCI energy surfaces were calculated using multiple active spaces, shown in Figures S9 and S10. The CASSCF surfaces for (7,8) and (7,8)' (shown in Figure S9a) are qualitatively similar, differing primarily in the energy differences between the well regions of DMP-D$_+$ and DMP-L$_+$. Interestingly, calculations with both active spaces give a minimum in the D$_+$ region that is off the diagonal, suggesting symmetry breaking in the wavefunction. The Mulliken spin populations (Table S5) for the lowest energy points (D1= -100°, D2=80°) show symmetry breaking and there is even some slight symmetry breaking for the symmetrical D1=-90, D2=90° points. Increasing the active space to (11,12) removes this symmetry breaking, giving a minimum at D1=-90, D2=90° and no symmetry-breaking in the Mulliken spin populations (Table S5). The slightly smaller active space of (11,11) reintroduces symmetry breaking, suggesting (11,12) to be the smallest active space to give a qualitatively correct energy surface.

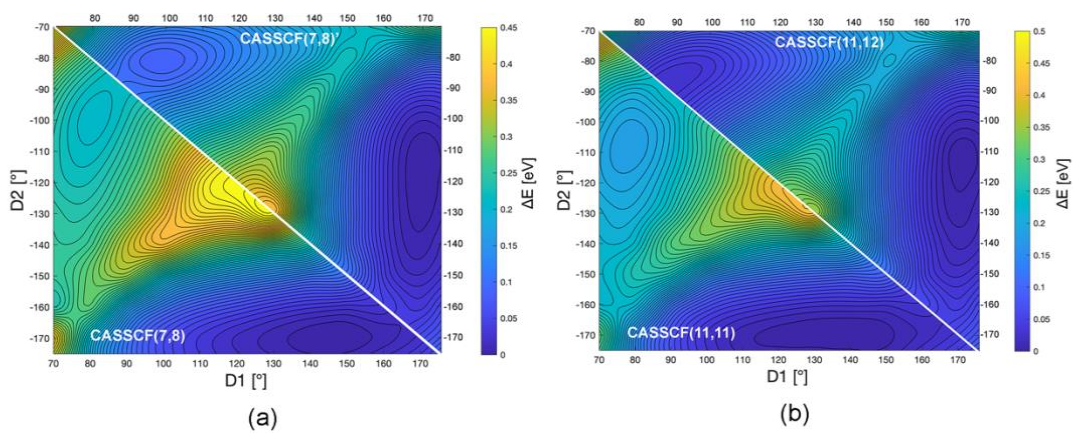

**Figure S9.** The potential energy surfaces calculated at the (a) CASSCF(7,8)'/aug-cc-pVDZ(upper) and CASSCF(7,8)/aug-cc-pVDZ(lower), and (b) CASSCF(11,12)/aug-cc-pVDZ(upper) and CASSCF(11,12)/aug-cc-pVDZ(lower) level of theory using the relaxed BHLYP structures.

The NEVPT2(7,8) and NEVPT2(7,8)' surfaces give a qualitatively correct energy surface, showing both L$_+$ and D$_+$ minima with no apparent symmetry breaking, with the delocalized state being overstabilized compared to experiment. As discussed in the manuscript and shown in Figure 2b, the L$_+$ minimum is absent from the NEVPT2(11,12) surface. Interestingly, removing a virtual orbital from the active space (CASSCF(11,11)) the localized minimum appears again at the NEVPT2 level of theory (Figure S10c). The CASSCF reference wavefunction in this case breaks symmetry for the D1=-80°, D2=80° structure ($s_{N1}$ =0.13, $s_{N2}$ = 0.69) as shown in Table S5. Finally, DMRG-NEVPT2(19,20) calculations show complete absence of the L$_+$ minimum. The NEVPT2 calculations are thus highly sensitive to the reference wavefunctions. The FIC-MRCI+Q calculations are interestingly not very sensitive to the active space as also seen in Figure S10 and in Figure 2b of the main article.

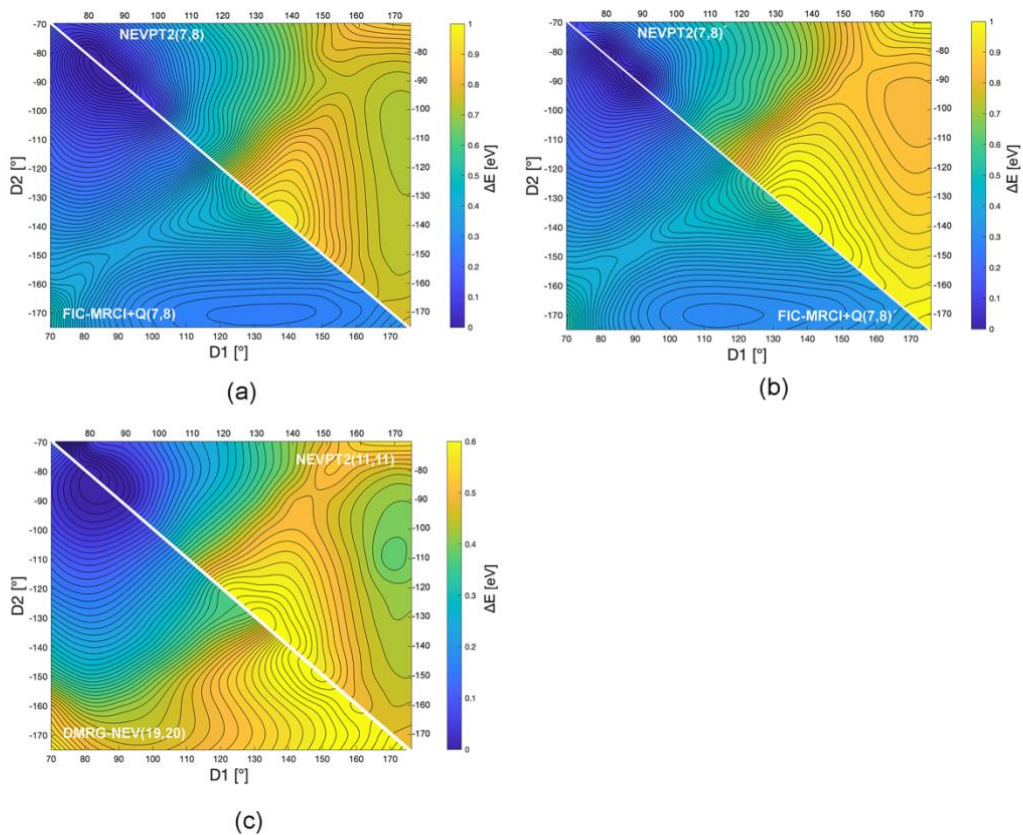

**Figure S10.** The potential energy surfaces calculated at the (a) NEVPT2(7,8)/aug-cc-pVDZ(upper) and MRCI(7,8)/cc-pVDZ (lower), (b) NEVPT2(7,8)'/aug-cc-pVDZ(upper) and MRCI(7,8)'/cc-pVDZ (lower) and (c) NEVPT2(11,11)/aug-cc-pVDZ(upper) and NEVPT2(19,20)/cc-pVDZ level of theory on the relaxed BHLYP structures.

## 6. Symmetry-breaking in the electronic structure of DMP-D+.

The DMP-D+ state is symmetric with respect to nuclear coordinates but the wavefunction can artificially break symmetry as can be seen by inspecting the difference in nitrogen spin populations. Table S5 compares Mulliken spin population on nitrogen atoms from different electronic structure methods for the DMP-D+ state.

**Table S5.** Mulliken spin populations on nitrogen atoms for the surface points which correspond to the DMP-D+ minimum found on the potential energy surfaces calculated at various levels of theory. BHLYP constraint-optimized structures were used.

|  | SN1 | SN2 |
|---|---|---|
| HF (D1=-80°, D2=80°), stable WF | 0.98 | 0.09 |
| HF (D1=-80°, D2=80°), unstable WF | 0.51 | 0.51 |
| MP2 (D1=-80°, D2=80°) relaxed density, stable HF WF | 0.31 | 0.44 |
| MP2 (D1=-80°, D2=80°) relaxed density, unstable HF WF | 0.28 | 0.28 |
| B3LYP (D1=-90°, D2=90°) | 0.40 | 0.40 |
| BHLYP (D1=-90°, D2=90°) | 0.43 | 0.43 |
| CCSD (D1=-90°, D2=90°) linearized, stable HF WF | 0.32 | 0.34 |
| CCSD (D1=-90°, D2=90°) unrelaxed, stable HF WF | 0.38 | 0.30 |
| CCSD (D1=-90°, D2=90°) linearized, unstable HF WF | 0.34 | 0.34 |
| CCSD (D1=-90°, D2=90°) unrelaxed, unstable HF WF | 0.34 | 0.34 |
| OO-CCD (D1=-90°, D2=90°) relaxed | 0.40 | 0.40 |
| CASSCF(3,2) (D1=-90°, D2=90°) | 0.03 | 0.79 |
| CASSCF(7,8) (D1=-90°, D2=90°) | 0.38 | 0.39 |
| CASSCF(7,8) (D1=-100°, D2=80°) | 0.62 | 0.18 |
| CASSCF(7,8)' (D1=-90°, D2=90°) | 0.36 | 0.37 |
| CASSCF(7,8)' (D1=-100°, D2=80°) | 0.63 | 0.13 |
| CASSCF(11,11) (D1=-80°, D2=80°) | 0.13 | 0.69 |
| CASSCF(11,12) (D1=-90°, D2=90°) | 0.37 | 0.37 |

## 7. Coupled-cluster calculations with different reference wavefunctions.

The CCSD and CCSD(T) PES surfaces were calculated using different reference wave functions: quasi-restricted orbitals (QRO), PBE orbitals and Brueckner orbitals. All calculations/orbital transformations started from a stable SCF solution. The data is presented in Figure S11. For CCSD, DMP-L$_+$ and DMP-D$_+$ minima were always present on the surfaces regardless of the reference wave function used, with only some changes in energies. The energy surface changed a bit compared to UHF-CCSD, especially regarding the position of the DMP-L+ minimum and the height of the saddlepoint region between DMP-L+ and of DMP-D$_+$. For CCSD(T), using QRO and PBE references did not influence the surface much (Figure S11a and S11b) compared to UHF-CCSD(T). However, the CCSD(T) PES obtained using Brueckner orbitals (Figure S11c) was found to be strangely chaotic in comparison to other CCSD(T) results. The reasons for this are not presently clear.

Calculations were also performed using orbital-optimized coupled-cluster theory where the orbitals are simultaneously relaxed at the CCSD level, resulting in optimal orbitals for the CCSD wavefunction, which results in the absence of single-excitations, hence the name OO-CCD. On top of the OO-CCD wavefunction the (T) correction was also calculated. The results for OO-CCD give a rather similar surface as UHF-CCSD, with some changes in the depth of the energy wells. The OO-CCD(T) surface suffers from the same problem as the other CCSD(T) surfaces.

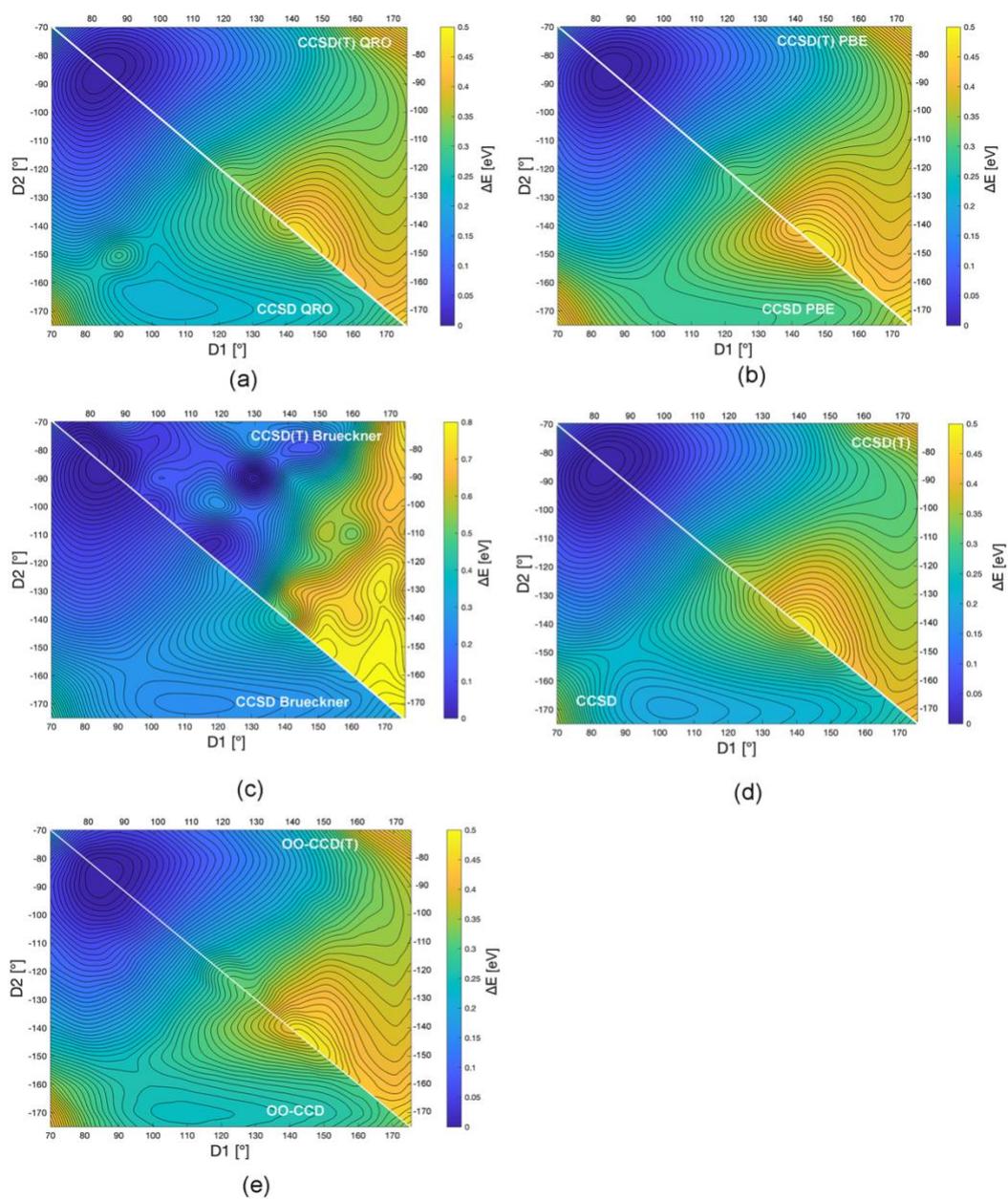

**Figure S11.** The potential energy surfaces calculated at the CCSD(T)/cc-pVDZ(upper) and CCSD/cc-pVDZ (lower) level of theory using different reference wave functions: (a) QRO, (b) PBE, (c) Brueckner and (d) UHF. Additionally, orbital optimized coupled cluster surfaces are shown in (e): OO-CCD(T) (upper) and OO-CCD (lower).

## 8. Fractional occupation density.

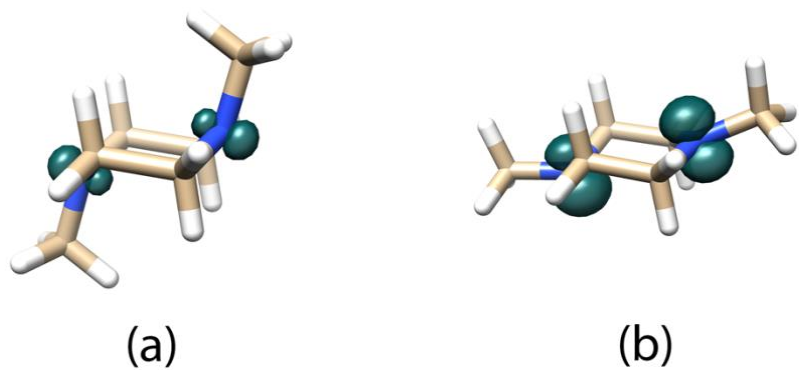

**Figure S12.** The fractional occupation density (FOD)[1] plotted for (a) DMP-D+ with FOD diagnostic = 0.17, and (b) DMP-D+ with FOD diagnostic = 0.61. The plots were done with isosurface level equal to 0.005. FOD calculations were performed at the default level of theory as implemented in ORCA: TPSS/def2-TZVP with electronic temperature of 5000 K.

(1) Grimme, S.; Hansen, A. A Practicable Real-Space Measure and Visualization of Static Electron-Correlation Effects. *Angew. Chem. Int. Ed.* **2015**, *54* (42), 12308–12313. https://doi.org/10.1002/anie.201501887.

## 9. Cartesian coordinates of optimized geometries.

Tables S6-17 show Cartesian coordinates of optimized structures at various levels of theory. A compressed directory with XYZ files for the entire 78-point BHLYP-constraint-optimized surface is provided as additional supporting information.

**Table S6.** The Cartesian coordinates [Å] of the DMP-D+ relaxed at the DMRG-CASSCF(19, 20)/aug-cc-pVDZ level.

```
N    0.212305   1.416373  -0.001500
C   -0.248340   0.785783   1.211312
C    0.254120  -0.784124   1.211257
N   -0.212302  -1.416377   0.001501
C    0.248344  -0.785789  -1.211308
C   -0.254120   0.784123  -1.211254
C    1.573277   2.025305  -0.005381
C   -1.573279  -2.025300   0.005376
H   -1.332585   0.778155   1.250034
H    0.165238   1.273758   2.085864
H   -0.155398  -1.270862   2.088414
H    1.338535  -0.776473   1.244889
H    0.155395   1.270860  -2.088413
H   -1.338536   0.776467  -1.244882
H    1.332589  -0.778154  -1.250029
H   -0.165233  -1.273763  -2.085860
H    1.676522   2.642331   0.880407
H    1.670270   2.644853  -0.890132
H    2.349074   1.257092  -0.009237
H   -1.670281  -2.644845   0.890128
H   -1.676524  -2.642330  -0.880410
H   -2.349072  -1.257082   0.009224
```

**Table S7.** The Cartesian coordinates [Å] of the DMP-L+ relaxed at the DMRG-CASSCF(19, 20)/aug-cc-pVDZ level.

```
N    0.294238   1.372261   0.061131
C   -0.239047   0.718941   1.266374
C    0.257460  -0.774084   1.294564
N   -0.184388  -1.402017   0.030465
C    0.283289  -0.750803  -1.211674
C   -0.215548   0.741639  -1.166234
C    0.085755   2.848443   0.072627
C   -1.340956  -2.343303   0.026647
H   -1.334120   0.737672   1.322099
H    0.162714   1.206743   2.148763
H   -0.164525  -1.322182   2.128548
H    1.341605  -0.789892   1.317156
H    0.203715   1.245192  -2.031464
H   -1.309201   0.760761  -1.243033
H    1.367726  -0.763208  -1.208824
H   -0.117397  -1.283416  -2.065483
H    0.552060   3.258262   0.963693
H   -0.975782   3.116119   0.064642
H    0.569359   3.274377  -0.801502
```

| | | | |
|---|---|---|---|
| H | -1.199248 | -3.056063 | 0.831935 |
| H | -1.381083 | -2.844700 | -0.932664 |
| H | -2.248464 | -1.764405 | 0.194099 |

**Table S8.** The Cartesian coordinates [Å] of the DMP-TS+ relaxed at the DMRG-CASSCF(19, 20)/aug-cc-pVDZ level.

| | | | |
|---|---|---|---|
| N | 0.192548 | 1.363640 | -0.001412 |
| C | -0.293297 | 0.755681 | 1.218973 |
| C | 0.259533 | -0.757763 | 1.265279 |
| N | -0.170676 | -1.400670 | 0.029946 |
| C | 0.305600 | -0.791062 | -1.205448 |
| C | -0.249212 | 0.722016 | -1.221012 |
| C | 0.529390 | 2.811449 | -0.017685 |
| C | -1.462834 | -2.142680 | 0.016961 |
| H | -1.382893 | 0.704917 | 1.283408 |
| H | 0.097461 | 1.284563 | 2.081288 |
| H | -0.153391 | -1.281495 | 2.119078 |
| H | 1.341143 | -0.723485 | 1.306599 |
| H | 0.173339 | 1.226641 | -2.082914 |
| H | -1.335572 | 0.668626 | -1.324075 |
| H | 1.388031 | -0.755547 | -1.206035 |
| H | -0.073257 | -1.339099 | -2.059768 |
| H | 1.096556 | 3.047028 | 0.877850 |
| H | -0.367182 | 3.432682 | -0.052606 |
| H | 1.143416 | 3.015696 | -0.889864 |
| H | -1.518691 | -2.755013 | 0.909661 |
| H | -1.494546 | -2.766603 | -0.868772 |
| H | -2.289670 | -1.431005 | 0.001409 |

**Table S9.** The Cartesian coordinates [Å] of the DMP-D+ relaxed at the CASSCF(11, 12)/aug-cc-pVDZ level.

| | | | |
|---|---|---|---|
| N | 0.202192 | 1.402171 | -0.001551 |
| C | -0.244364 | 0.790500 | 1.189406 |
| C | 0.249789 | -0.788688 | 1.189543 |
| N | -0.202193 | -1.402170 | 0.001551 |
| C | 0.244364 | -0.790500 | -1.189406 |
| C | -0.249789 | 0.788689 | -1.189542 |
| C | 1.562955 | 2.014476 | -0.005436 |
| C | -1.562955 | -2.014477 | 0.005436 |
| H | -1.329389 | 0.777858 | 1.239322 |
| H | 0.167629 | 1.270278 | 2.069963 |
| H | -0.158326 | -1.267085 | 2.072666 |
| H | 1.335025 | -0.776006 | 1.234612 |
| H | 0.158325 | 1.267085 | -2.072666 |
| H | -1.335025 | 0.776007 | -1.234613 |
| H | 1.329388 | -0.777858 | -1.239322 |
| H | -0.167629 | -1.270278 | -2.069963 |
| H | 1.666875 | 2.630768 | 0.880928 |
| H | 1.659829 | 2.634881 | -0.889754 |
| H | 2.341014 | 1.247893 | -0.010352 |
| H | -1.659828 | -2.634882 | 0.889754 |
| H | -1.666874 | -2.630769 | -0.880928 |
| H | -2.341014 | -1.247895 | 0.010352 |

**Table S10.** The Cartesian coordinates [Å] of the DMP-L+ relaxed at the CASSCF(11, 12)/aug-cc-pVDZ level.

| | | | |
|---|---|---|---|
| N | 0.303770 | 1.381037 | 0.061558 |
| C | -0.239817 | 0.762230 | 1.245709 |
| C | 0.172436 | -0.747042 | 1.279992 |
| N | -0.244290 | -1.368873 | 0.030163 |
| C | 0.193413 | -0.722977 | -1.199937 |
| C | -0.220271 | 0.785570 | -1.143221 |
| C | 0.144702 | 2.860946 | 0.074684 |
| C | -1.266639 | -2.450651 | 0.025358 |
| H | -1.334510 | 0.836443 | 1.306823 |
| H | 0.177305 | 1.224630 | 2.135820 |
| H | -0.286948 | -1.269482 | 2.111190 |
| H | 1.255724 | -0.821619 | 1.335077 |
| H | 0.211633 | 1.264929 | -2.017177 |
| H | -1.313742 | 0.860877 | -1.221003 |
| H | 1.277732 | -0.794643 | -1.235894 |
| H | -0.249339 | -1.228998 | -2.049685 |
| H | 0.623494 | 3.258703 | 0.964929 |
| H | -0.908813 | 3.161968 | 0.068829 |
| H | 0.638243 | 3.275972 | -0.799477 |
| H | -1.059461 | -3.122107 | 0.851865 |
| H | -1.222844 | -2.974301 | -0.922145 |
| H | -2.243618 | -1.986274 | 0.158407 |

**Table S11.** The Cartesian coordinates [Å] of the DMP-L+ relaxed at the CASSCF(7, 8)'/aug-cc-pVDZ level.

| | | | |
|---|---|---|---|
| N | 0.292527 | 1.378592 | 0.061359 |
| C | -0.245198 | 0.745783 | 1.241929 |
| C | 0.193263 | -0.752082 | 1.283010 |
| N | -0.197633 | -1.394865 | 0.031850 |
| C | 0.220698 | -0.731091 | -1.199666 |
| C | -0.216941 | 0.766847 | -1.142699 |
| C | 0.125293 | 2.826852 | 0.072040 |
| C | -1.101067 | -2.537398 | 0.038268 |
| H | -1.341422 | 0.802172 | 1.294988 |
| H | 0.156330 | 1.217358 | 2.134611 |
| H | -0.264140 | -1.281783 | 2.111221 |
| H | 1.277451 | -0.809971 | 1.349005 |
| H | 0.207210 | 1.252886 | -2.017007 |
| H | -1.311319 | 0.825698 | -1.221574 |
| H | 1.306236 | -0.788680 | -1.240457 |
| H | -0.216966 | -1.244816 | -2.047603 |
| H | 0.594662 | 3.238450 | 0.961752 |
| H | -0.929553 | 3.126126 | 0.061556 |
| H | 0.616523 | 3.253976 | -0.798341 |
| H | -0.765772 | -3.240594 | 0.795356 |
| H | -1.111068 | -3.003308 | -0.939708 |
| H | -2.098976 | -2.180812 | 0.295258 |

**Table S12.** The Cartesian coordinates [Å] of the DMP-L+ relaxed at the CASSCF(7, 8)/aug-cc-pVDZ level.

| | | | |
|---|---|---|---|
| N | 0.319155 | 1.356402 | 0.054090 |
| C | -0.237353 | 0.727931 | 1.230812 |
| C | 0.172779 | -0.740212 | 1.278152 |
| N | -0.094762 | -1.406117 | -0.000265 |
| C | 0.167954 | -0.691352 | -1.253197 |
| C | -0.239205 | 0.774738 | -1.145538 |
| C | 0.158750 | 2.834810 | 0.083068 |
| C | -0.444655 | -2.852050 | -0.008165 |
| H | -1.332646 | 0.806329 | 1.269288 |
| H | 0.155375 | 1.202237 | 2.126115 |
| H | -0.350540 | -1.287286 | 2.056348 |
| H | 1.248104 | -0.832561 | 1.435650 |
| H | 0.152548 | 1.283566 | -2.022125 |
| H | -1.334465 | 0.855526 | -1.179257 |
| H | 1.242274 | -0.780315 | -1.420235 |
| H | -0.361516 | -1.204487 | -2.049540 |
| H | 0.644822 | 3.224842 | 0.972827 |
| H | -0.895516 | 3.133828 | 0.089686 |
| H | 0.643466 | 3.259229 | -0.791564 |
| H | 0.127522 | -3.348705 | 0.769583 |
| H | -0.221085 | -3.269217 | -0.982906 |
| H | -1.509443 | -2.932800 | 0.206237 |

**Table S13.** The Cartesian coordinates [Å] of the DMP-D+ relaxed at the BHLYP/aug-cc-pVDZ level.

| | | | |
|---|---|---|---|
| N | 0.22030638483835 | 1.38335169439629 | -0.00141428270678 |
| C | -0.23307706173303 | 0.76534307809786 | 1.18740587238139 |
| C | 0.23894425174175 | -0.76386646452091 | 1.18722001517577 |
| N | -0.22030684419671 | -1.38335113899910 | 0.00141431037829 |
| C | 0.23307659981281 | -0.76534262310306 | -1.18740587001465 |
| C | -0.23894470846030 | 0.76386687853965 | -1.18721999841372 |
| C | 1.53471993591523 | 1.99276339527823 | -0.00517159621707 |
| C | -1.53471981040781 | -1.99276394447674 | 0.00517158067032 |
| H | -1.32174960069968 | 0.77017282021861 | 1.23316844148976 |
| H | 0.17448322791798 | 1.25447472839007 | 2.06793203095279 |
| H | -0.16433157233165 | -1.25187750268697 | 2.07034182284149 |
| H | 1.32782459474641 | -0.76866820858800 | 1.22766333849882 |
| H | 0.16433122635944 | 1.25187782151717 | -2.07034182920387 |
| H | -1.32782506498139 | 0.76866867233390 | -1.22766348128770 |
| H | 1.32174917170525 | -0.77017245222663 | -1.23316826841778 |
| H | -0.17448366508040 | -1.25447438095232 | -2.06793194920358 |
| H | 1.64798713622730 | 2.61399603089722 | 0.87961502815664 |
| H | 1.64232763481302 | 2.61519954488115 | -0.88983629749963 |
| H | 2.33102223344338 | 1.23797485970780 | -0.00826747341729 |
| H | -1.64232681989818 | -2.61520057736802 | 0.88983602393025 |
| H | -1.64798657934418 | -2.61399616615818 | -0.87961535284821 |
| H | -2.33102267038759 | -1.23797606517800 | 0.00826793475475 |

**Table S14.** The Cartesian coordinates [Å] of the DMP-L+ relaxed at the BHLYP/aug-cc-pVDZ level.

| | | | |
|---|---|---|---|
| N | -1.378932 | -0.003352 | -0.226901 |
| C | -0.705445 | 1.187327 | 0.222762 |
| C | 0.700124 | 1.248790 | -0.380457 |
| N | 1.410727 | 0.013527 | -0.128835 |
| C | 0.713221 | -1.228795 | -0.388546 |
| C | -0.690703 | -1.187437 | 0.218188 |
| C | -2.794842 | -0.012378 | 0.105594 |
| C | 2.757891 | -0.005867 | 0.378014 |
| H | -0.629913 | 1.247429 | 1.319155 |
| H | -1.245791 | 2.068554 | -0.118935 |
| H | 1.280830 | 2.076787 | 0.016026 |
| H | 0.618690 | 1.350018 | -1.467284 |
| H | -1.221707 | -2.073941 | -0.124470 |
| H | -0.612504 | -1.249663 | 1.314463 |
| H | 0.632766 | -1.324689 | -1.475509 |
| H | 1.306062 | -2.051484 | 0.002621 |
| H | -3.272092 | 0.866545 | -0.323696 |
| H | -2.973101 | -0.015002 | 1.188882 |
| H | -3.261381 | -0.895772 | -0.326273 |
| H | 3.335898 | -0.751465 | -0.168224 |
| H | 2.728250 | -0.305096 | 1.431918 |
| H | 3.209944 | 0.976712 | 0.288157 |

**Table S15.** The Cartesian coordinates [Å] of the DMP-TS+ relaxed at the BHLYP/aug-cc-pVDZ level.

```
 N    1.131597   0.828767   0.002660
 C    0.422674   0.572147   1.208039
 C   -0.103337  -0.901559   1.172916
 N   -0.862151  -1.059246  -0.030418
 C   -0.164942  -0.794520  -1.251810
 C    0.363590   0.677288  -1.184182
 C    2.197429   1.811730   0.019502
 C   -2.298031  -0.935325   0.015275
 H   -0.446649   1.226865   1.359272
 H    1.085291   0.678073   2.064392
 H   -0.726512  -1.109469   2.037769
 H    0.754386  -1.570691   1.154351
 H    0.984504   0.857744  -2.059074
 H   -0.510080   1.341604  -1.235907
 H    0.690230  -1.461981  -1.335208
 H   -0.831287  -0.926148  -2.099190
 H    1.819671   2.838707   0.074090
 H    2.795376   1.708868  -0.884254
 H    2.839693   1.630581   0.879473
 H   -2.684936  -1.469849   0.878986
 H   -2.731732  -1.332189  -0.898356
 H   -2.570440   0.124572   0.106291
```

**Table S16.** The Cartesian coordinates [Å] of the DMP-L+ relaxed at the HF/aug-cc-pVDZ.

```
 N    0.232020   1.385041   0.000000
 C   -0.232020   0.774264   1.186092
 C    0.232020  -0.774264   1.186092
 N   -0.232020  -1.385041   0.000000
 C    0.232020  -0.774264  -1.186092
 C   -0.232020   0.774264  -1.186092
 C    1.558369   1.986895   0.000000
 C   -1.558369  -1.986895   0.000000
 H   -1.317991   0.785464   1.230631
 H    0.177979   1.252343   2.069046
 H   -0.177979  -1.252343   2.069046
 H    1.317991  -0.785464   1.230631
 H    0.177979   1.252343  -2.069046
 H   -1.317991   0.785464  -1.230631
 H    1.317991  -0.785464  -1.230631
 H   -0.177979  -1.252343  -2.069046
 H    1.670110   2.607465   0.883069
 H    1.670110   2.607465  -0.883069
 H    2.344777   1.225810   0.000000
 H   -1.670110  -2.607465   0.883069
 H   -1.670110  -2.607465  -0.883069
 H   -2.344777  -1.225810   0.000000
```

**Table S17.** The Cartesian coordinates [Å] of the DMP-L1+ relaxed at the HF/aug-cc-pVDZ level.

```
 N    0.227141   1.382003  -0.035556
 C   -0.189598   0.740863   1.211502
 C    0.217068  -0.752495   1.168054
 N   -0.238065  -1.427470  -0.020330
 C    0.228219  -0.767156  -1.212665
 C   -0.178411   0.724690  -1.278116
 C    1.085673   2.557637  -0.067763
```

```
C   -1.640521  -1.817144  -0.024591
H   -1.272369   0.848436   1.270114
H    0.266342   1.262446   2.045722
H   -0.184361  -1.232824   2.055940
H    1.303838  -0.817897   1.224772
H    0.288190   1.236666  -2.113286
H   -1.259990   0.833737  -1.350983
H    1.315619  -0.832869  -1.258030
H   -0.164288  -1.259115  -2.098163
H    1.113945   3.021244   0.910892
H    0.699157   3.250746  -0.810291
H    2.086887   2.240090  -0.363070
H   -1.845805  -2.418953   0.856988
H   -1.836885  -2.431845  -0.899302
H   -2.346606  -0.977184  -0.034374
```